\documentclass[iop]{emulateapj}
\usepackage{amssymb,amsmath}
\usepackage[colorlinks=true,linkcolor=blue,anchorcolor=blue,citecolor=blue,urlcolor=blue]{hyperref}
\usepackage{ulem}
\usepackage{color}
\usepackage{natbib}

\slugcomment{}

\shorttitle{Quiescent Compact Galaxies at Intermediate Redshift in the COSMOS Field. The Number Density}
\shortauthors{Damjanov et al.}

\begin{document}


\title{Quiescent Compact Galaxies at Intermediate Redshift in the COSMOS Field. The Number Density}


\author{Ivana Damjanov\altaffilmark{1}, Margaret J. Geller\altaffilmark{2}, H. Jabran Zahid\altaffilmark{2}, Ho Seong Hwang\altaffilmark{2,3}}
\altaffiltext{1}{Harvard-Smithsonian Center for Astrophysics, 60 Garden Street, Cambridge, MA 02138}
\altaffiltext{2}{Smithsonian Astrophysical Observatory, 60 Garden St., Cambridge, MA 02138}
\altaffiltext{3}{School of Physics, Korea Institute for Advanced Study, 85 Hoegiro, Dongdaemun-gu, Seoul 130-722, Republic of Korea}


\begin{abstract}

We investigate the evolution of compact galaxy number density over the redshift range $0.2<z<0.8$. Our sample consists of galaxies with secure spectroscopic redshifts observed in the COSMOS field. With the large uncertainties, the compact galaxy number density trend with redshift is consistent with a constant value over the interval $0.2<z<0.8$. Our number density estimates are similar to the estimates at $z>1$ for equivalently selected compact samples. Small variations in the abundance of the COSMOS compact sources as a function of redshift correspond to known structures in the field. The constancy of the compact galaxy number density is robust and insensitive to the compactness threshold or the stellar mass range (for $M_\ast>10^{10}\, M_\odot$). To maintain constant number density any size growth of high-redshift compact systems with decreasing redshift must be balanced by formation of quiescent compact systems at $z<1$.    

\end{abstract}



\section{Introduction}

Structural evolution of quiescent galaxies provides important insights into the mass assembly of the most massive galaxies in the local universe. A suite of studies have demonstrated that these massive quiescent galaxies at $z>1$ are several times smaller than typical similarly massive passive systems at $z\sim0$ \citep[e.g.,][]{Daddi2005, Longhetti2007,Trujillo2007, Toft2007, Zirm2007, Cimatti2008, vanDokkum2008, Buitrago2008, Damjanov2009, Damjanov2011, McLure2012, vandeSande2013, Belli2014}. Potential sources of systematic errors in both galaxy size measurements (based on the surface brightness profile fitting) and galaxy stellar mass estimates do not significantly affect these findings \citep{Muzzin2009, Szomoru2012}. 

Evolution in the abundance of compact systems with cosmic time remains an open issue, particularly at $0<z<1$. At high redshift  ($z>1$),  the number density of the compact quiescent sources is in the range $10^{-5}-10^{-4}\, \mathrm{Mpc}^{-3}$ \citep[e.g.,][]{Bezanson2009, Saracco2010a, Cassata2011, Cassata2013, Stefanon2013, Barro2013, vanderWel2014}. The exact value of the number density depends on the definition of compactness and on the minimal stellar mass probed in the study.  

In the local universe, results vary dramatically among surveys. Based on the Sloan Digital Sky Survey (SDSS), the number of massive compact systems at $z\sim0$ is negligible; the number density is more than three orders of magnitude below the high redshift values \citep{Trujillo2009, Taylor2010}. The apparent disappearance of the massive compact systems from $z>1$ to $z=0$ requires a set of
dry mergers or alternative exotic mechanisms to drive galaxy size growth \citep[e.g.,][]{Nipoti2009a,Nipoti2009b, Oogi2013,Belli2014a}. However, the apparent sharp decline in compact galaxy number density may be the result of biased galaxy selection and poor imaging quality of the SDSS \citep{Taylor2010, Carollo2013}. 

In contrast with the SDSS-based studies, results from the PM2GC survey at $z<0.1$ \citep{Calvi2011} suggest that the abundance of local compact systems is comparable with the number density at high redshift \citep{Poggianti2013, Poggianti2013a}. The WINGS survey of nearby clusters \citep{Fasano2006} shows high fractions of compact systems in galaxy cluster environments \citep{Valentinuzzi2010a}, suggesting a link between compact massive systems and the galaxy local density. 

Using a sample of point-source objects from the SDSS photometric database with signatures of passive evolution in their Baryon Oscillation Spectroscopic Survey spectra \citep[BOSS,][]{Dawson2013} we identified a large sample of intermediate-redshift massive compact galaxies \citep{Damjanov2013, Damjanov2014}. These objects are at least two times smaller than their massive $z\sim0$ SDSS analogs. Hard lower limits on the number densities of these systems at $0.2<z<0.6$ clearly show that the abundance of massive compact systems does not decline steeply in this redshift range \citep{Damjanov2014}.

The intermediate redshift regime is an important link between compact samples in the nearby and high-redshift universe. The range $0.2<z<0.8$ presents a sweet spot for identifying the compact quiescent galaxy population and for probing their structural, dynamical and environmental properties. The volumes surveyed in this redshift range are large and compact massive systems are bright enough to be targeted by large spectroscopic surveys. Spectro-photometric studies of the intermediate-redshift compact population can further constrain the mechanisms governing the compact massive galaxy formation and evolution. 

Here we examine a sample of massive ($M_\ast>10^{10}\, M_\odot$) quiescent galaxies in the COSMOS field with available size measurements based on surface brightness profile fitting and with known spectroscopic redshifts to measure the compact galaxy number density at $0.2<z<0.8$. In Section~\ref{MCGCosmos} we introduce the parent sample of quiescent galaxies and describe the selection functions we use to extract compact galaxy samples.  In Section~\ref{abundanceexample} we select one compact sample to demonstrate the method we employ to correct for observational selection effects and to estimate the compact galaxy number density. In Section~\ref{dis} we examine the effects that the cosmic variance and compact galaxy definition have on derived intermediate-redshift number densities. In an accompanying paper we investigate the Fundamental Plane defined by the most massive intermediate-redshift compact systems in the COSMOS field \citep[][, hereafter Paper II]{Zahid2015}. We adopt a cosmological model with $\Omega_{\Lambda}=0.7$, $\Omega_{M} = 0.3$, and $H_0 = 70$~km~s$^{-1}$~Mpc$^{-1}$.  

\section{Identifying Massive Compact Galaxies in the COSMOS field}\label{MCGCosmos}

The public data for the COSMOS field offer a unique opportunity to identify massive compact galaxies and to explore the evolution in their number density as a function of redshift. The available datasets include: 1) the catalog of structural parameters for objects in the 1.64 deg$^2$ field covered by the HST ACS F814W imaging \citep{Scarlata2007, Sargent2007, Koekemoer2007}\footnote{\url{http://irsa.ipac.caltech.edu/data/COSMOS/tables/morphology/cosmos_morph_zurich_1.0.tbl}}, 2) photometric catalog with stellar masses for galaxies in the 1.5~deg$^2$ (unmasked) field of the UltraVISTA survey \citep{Ilbert2013, McCracken2012}\footnote{\url{http://vizier.cfa.harvard.edu/viz-bin/VizieR?-source=J/A+A/556/A55}}, and 3) a compilation of available spectroscopic redshifts in the COSMOS field \citep{Davies2014}\footnote{\url{http://ict.icrar.org/cutout/G10/panchromaticDR.php}}.  

Here we describe our approach to constructing a catalog of compact massive galaxies in the redshift range 0.2$ < z < $ 0.8. Because the identification may depend on the approach to size measurement, we remeasure sizes for a subset of the smallest objects. We then extract catalogs based on the various definitions of compactness and we summarize the samples.

\subsection {Sample Selection}\label{COSMOSsample}

\begin{figure*}
\begin{centering}
\includegraphics[scale=0.35]{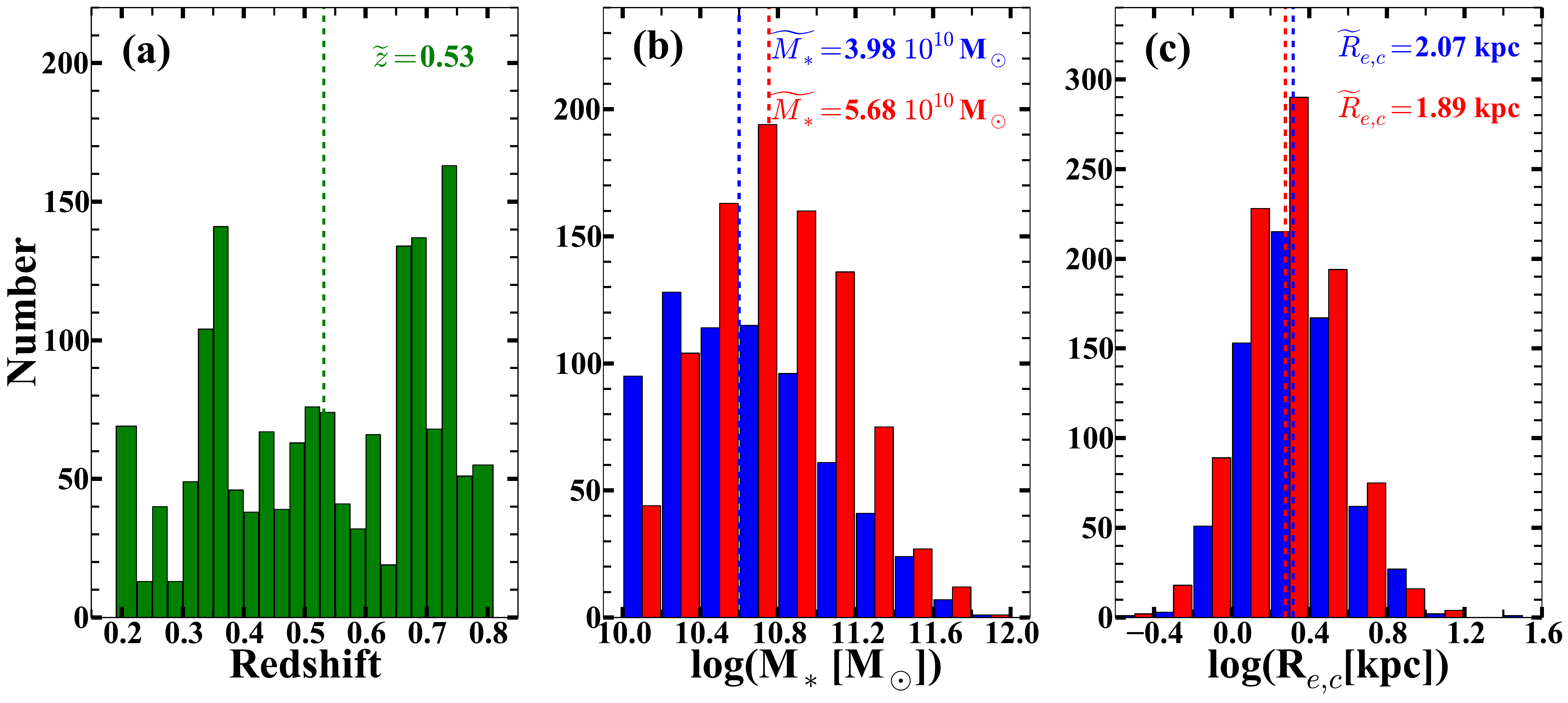}
\caption{The distribution of spectroscopic redshifts (a), stellar masses (b) and circularized effective radii (c) for the intermediate redshift massive quiescent galaxies selected in COSMOS field. Dashed lines indicate median values for each quantity. The stellar mass and effective radius panels show low redshift ($0.2\leqslant z<0.5$; blue) and high redshift ($0.5\leqslant z\leqslant 0.8$; red) subsamples. \label{f1}}
\end{centering}
\end{figure*}

We select the parent quiescent sample using the following criteria: 
\begin{itemize}

\item Galaxies are in the redshift range $0.2\leqslant z\leqslant 0.8$; 

\item Galaxy stellar masses are $M_\ast>10^{10}\, M_\sun$; 

\item Galaxies are defined as passive: $(NUV -r)>3\times(r-J)+1$ and $(NUV-r)> 3.1$ \citep{Ilbert2010, Ilbert2013}.  

\end{itemize}

\noindent In addition, we select only intermediate redshift quiescent galaxies with robust spectroscopic redshifts reported in the \citet{Davies2014} redshift catalog.  These objects include: 

\begin{itemize}

\item zCOSMOS-observed sources with redshifts classified as 99\% reliable \citep[based on the classification scheme described in][]{Lilly2009}, 

\item Targets with AUTOZ redshifts that are confirmed by PRIsm MUlti-Object Survey \citep[PRIMUS,][]{Coil2011, Cool2013} or zCOSMOS surveys (parameter Z\_GEN$=1,\, 2$ in the \citealt{Davies2014} catalog),

\item VIMOS VLT Deep Survey \citep{leFevre2005} targets, 

\item SDSS DR10 \citep{Ahn2014} spectroscopic sources. 

\end{itemize}

\noindent We crossmatch this sample with the COSMOS catalog of galaxy morphology based on Gim2D \citep{Simard2002} fitting, using the cutoff magnitude $I_\mathrm{UVISTA}(AB)=23$~mag to include only systems with reliable size measurements \citep{Sargent2007}. The final parent sample contains 1599 massive quiescent galaxies with available size measurements and with spectroscopic  redshifts in the range $0.2\leqslant z\leqslant 0.8$. 

Figure~\ref{f1} shows the distribution of the COSMOS parent galaxy sample properties. The spectroscopic redshift distribution in Figure~\ref{f1}a shows prominent peaks in galaxy number counts at $z\sim0.35$, $0.65<z<0.7$, and $z\sim0.725$, where there are significant galaxy overdensities  \citep[e.g.,][]{Scoville2007, Guzzo2007, Kovac2010, Masters2011, Scoville2013}. 

\subsection {Stellar Mass Estimates}\label{masses}

We adopt the stellar masses given in the \citet{Ilbert2013} cataolog. \citet{Ilbert2013} estimate the stellar masses for COSMOS galaxies by comparing stellar population synthesis models to 30 bands of ultraviolet to infrared photometry. The multi-band SEDs are fit using the LePHARE\footnote{\url{http://www.cfht.hawaii.edu/~arnouts/LEPHARE/lephare.html}} code written by Arnouts \& Ilbert \citep{Arnouts1999, Ilbert2006}. A library of synthetic spectra are generated based on the \citet{Bruzual2003} stellar population models. Synthetic SEDs are reddened using the \citet{Calzetti2000} extinction law with a range of $E(B-V)$ values. Exponential star formation histories are adopted with $\tau = 0 - 30$ Gyr and models span three metallicities ($[0.2, 0.4, 1] \times Z_\odot$). The stellar mass is the scale factor that minimizes the difference between the synthetic and observed SEDs. The median of the stellar mass distribution is the stellar mass given in the \citet{Ilbert2013} catalog. For consistency, we use the LePHARE code with similar parameters to calculate the stellar masses of SDSS galaxies in Section~\ref{compprop}.

Figure \ref{f1}b shows the histogram of the stellar mass distribution for the parent sample. The blue and red histograms show the mass distribution for galaxies with redshifts below and above the median redshift ($z\sim0.5$), respectively. The galaxies differ in the expected sense that low mass galaxies are relatively less abundant in the higher redshift sample. In general the decline in the mass distribution below the characteristic mass (where the mass function of quiescent population itself stays relatively constant, Ilbert et al. 2013) reflects observational selection that we must account for in calculating the total number density of objects as a function of redshift (see Section~\ref{obseff}). 

\subsection {Size Measurements}\label{sizes}

\begin{figure*}
\begin{centering}
\hspace*{-0.35in}
\includegraphics[scale=0.35]{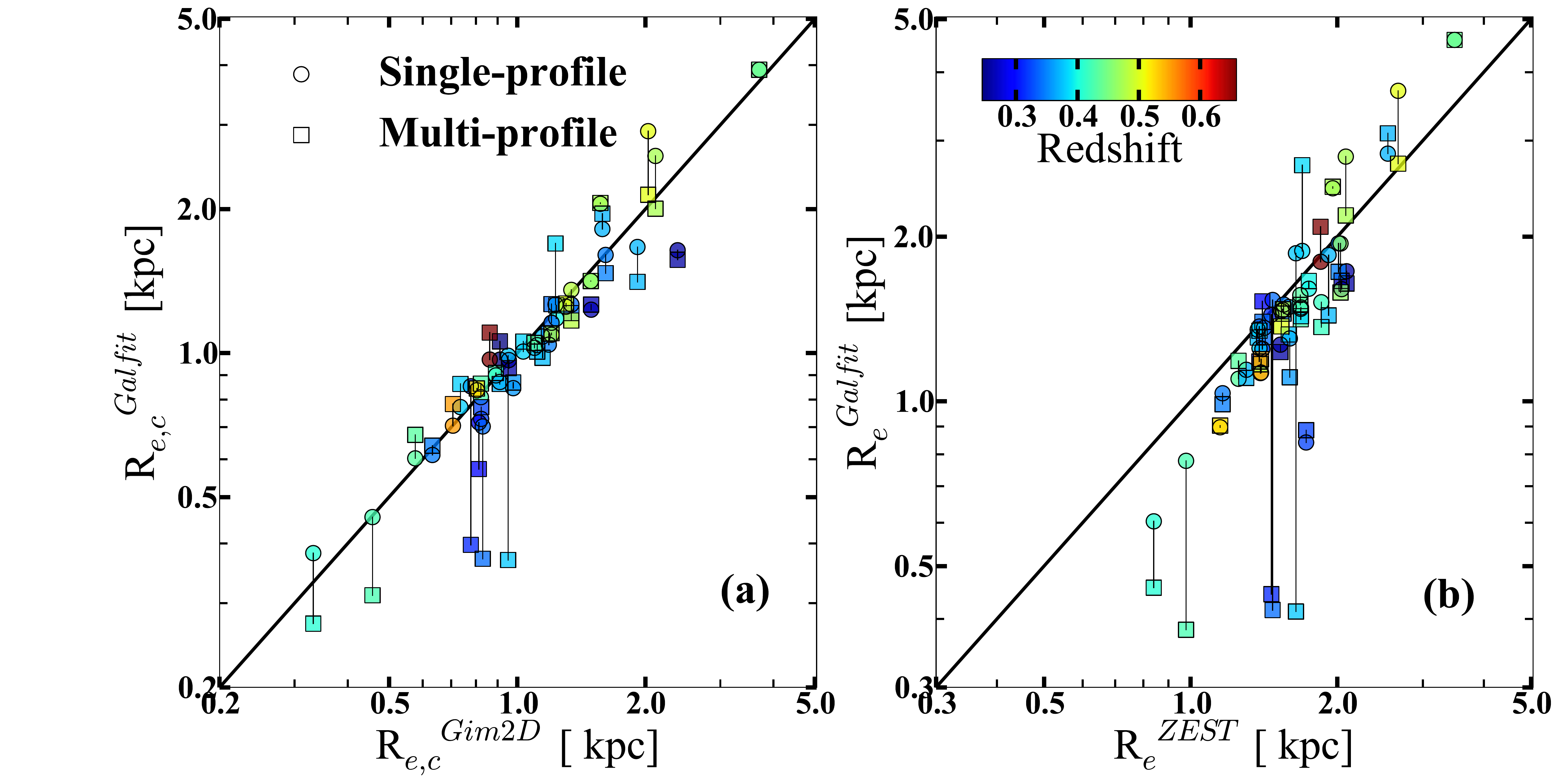}
\caption{(a):  Comparison between circularized effective radii of the best fit Galfit and Gim2D models for 41 intermediate redshift compact galaxies in the COSMOS. (b): Effective radius along the major axis of the best-fit Galfit model versus the $ZEST+$ semi-major axis length of the ellipse encompassing 50\% of total light. In both panels  circles correspond to two single \citet{SersicJoseLuis1968} profile models. The $y-$coordinate for squares is the effective radius of the best-fit multi-S\'ersic profile based on Galfit. All points are color-coded by redshift. \label{f2}}  
\end{centering}
\end{figure*}

Robustness of the size measurements is a crucial element in understanding the nature of compact massive galaxies and their connection to the entire quiescent population. Next we check that the sizes of some of the most compact objects in the COSMOS field are insensitive to the approach to size determination.

To test size estimates of \citet{Sargent2007}, we choose 41 quiescent COSMOS galaxies with measured sizes and stellar masses, but which are classified as point sources in the SDSS DR10 photometric catalog. We include only the brightest objects with $r_\mathrm{SDSS}<21.8$ to match the selection criteria we previously used to estimate the lower limit on the number density of intermediate redshift compact systems in BOSS \citep{Damjanov2014}. 

Using Galfit software v3.0.4 \citep{Peng2010}, we fit the HST ACS F814W images of SDSS/COSMOS targets with two sets of two-dimensional models: a single S\'ersic profile and a composite model that may include up to three S\'ersic profiles. The functional form of the single-profile surface brightness model is:

\begin{equation} 
\Sigma(r)=\Sigma_e\times \mathrm{exp} \left[-\kappa\left(\left(\frac{r}{R_e}\right)^{\frac{1}{n}}-1\right)\right], \label{eq:sersic}
\end{equation}
 
\noindent where $R_e$ is the effective radius encompassing half of the total flux, $\Sigma_e$ is the surface brightness at $R_e$, $n$ is S\'{e}rsic index describing the central concentration, and $\kappa(n)$ is a normalization factor. For each galaxy we select several ($2-5$) bright stars within a $45\arcsec$ radius and use IRAF {\it daophot} routines to construct corresponding PSF. Galfit convolves this PSF with a S\'ersic surface brightness profile and uses the Levenberg-Marquardt algorithm to match this analytic model to the observed surface brightness profile. We compare the resulting best-fit circularized effective radius $R_{e,c}^{\mathrm{Galfit}}=R_{e}\sqrt{q}$ (where $q$ is the axial ratio of the best-fit model) to the same parameter of the best-fit Gim2D model, $R_{e,c}^{\mathrm{Gim2D}}$. The two quantities are in excellent agreement (circles in Figure~\ref{f2}a), with $R_{e,c}^{\mathrm{Gim2D}}/R_{e,c}^{\mathrm{Galfit}}=1.00\pm0.13$. Thus the two methods based on the comparison between the S\'ersic model and observed surface brightness profiles, Gim2D and Galfit,  provide almost identical size estimates.  

We compare the effective radius along the major axis of the best-fit single-S\'ersic profile with the effective radius within the elliptical aperture \citep[measured using the $ZEST+$ algorithm, ][]{Scarlata2007} for selected objects in Figure~\ref{f2}b. These two quantities also agree very well: $R_e^{\mathrm{Galfit}}/R_e^{ZEST+}=0.94\pm0.17$. However, in the regime of high surface brightness and compact sizes, Galfit measurements (or similar techniques based on matching with a PSF-convolved model) are more reliable for recovering intrinsic sizes of simulated galaxies \citep{Carollo2013}.  

Because the HST ACS images reveal a bulge+disk structure for most of these targets, we also fit the observed surface brightness profiles of compact massive galaxies in our COSMOS/SDSS sample with a composite models. Indeed, only three out of 41 objects are well-fitted with a single bulge. After obtaining the best-fit multi-S\'ersic profile for each target, we use the {\it ellipse} task in the IRAF stsdas.analysis.isophote package to extract the intrinsic one-dimensional profile (i.e., before the PSF convolution) and to perform a curve of growth analysis. 

The resulting multi-profile effective radius is on average very close to the single-profile model size: the mean value of the ratio $R_{e,c}^{\mathrm{multi-Galfit}}/R_{e,c}^{\mathrm{single-Galfit}}=0.95$ (squares in Figure~\ref{f2}a). However, 10(eight) out of 41 objects exhibit multi-profile sizes that are more than $10\%$($20\%$) smaller than their single-S\'ersic model sizes. The composite structure of these galaxies includes a faint disk component surrounding a prominent highly concentrated bulge which drives the intrinsic effective radius of the multi-S\'ersic profile down to 37\% of the single-S\'ersic profile size (for the most extreme case). In contrast, only 5(1) objects have multi-profile models which are $10\%$($20\%$) more extended than their single-S\'ersic fits. The ratio between the multi-profile Galfit effective radius and the $ZEST+$ size shows a similar trend (squares in Figure~\ref{f2}b). These observed trends confirm the results based on size measurements of simulated multi-component galaxies: modeling such systems with a single S\'ersic component produces slightly overestimated sizes \citep{Davari2014}. Although almost all of the sample galaxies show a disk component in addition to a prominent bulge, their composite structure is at least  as compact as their best-fit single-profile model.

Figure~\ref{f1}c shows the size distribution of the massive quiescent galaxies. They span the range  $0.3\, \mathrm{kpc} \lesssim R_{e,c} \lesssim 27$~kpc with a median value of $\widetilde{R}_{e,c}=2$~kpc. The size distributions above and below the median redshift are more similar than the mass distributions. This similarity suggests that correction for an observational bias related to size measurements in this redshift range is unnecessary.

\subsection{The Compact Galaxy Catalogs}\label{compactsamples}

We construct several samples of $0.2<z<0.8$ compact galaxies using different compactness criteria defined in the literature. All of these definitions are based on the position of selected objects in the effective radius vs. stellar mass diagram. They account for the positive trend that galaxy sizes exhibit as a function of stellar mass. Table~\ref{tab1} summarizes the properties of each selected sample.

\begin{deluxetable*}{ccccccccc}
\tabletypesize{\scriptsize}
\setlength{\tabcolsep}{1pt} 
\tablecaption{The properties of compact COSMOS samples \label{tab1}}
\tablewidth{0pt}
\tablehead{ \colhead{Definition} & \colhead{Reference} & \colhead{Sample size\tablenotemark{a}} & \colhead{$\langle \log\left(\frac{M_\ast}{M_\sun}\right)\rangle$} & \colhead{$\sigma_{\log M_\ast}$} & \colhead{$\langle R_e\rangle$\tablenotemark{b}}& \colhead{$\sigma_{R_e}$} & \colhead{$z_{min}$} &  \colhead{$z_{max}$} \\
\colhead{} & \colhead{} & \colhead{} & \colhead{} & \colhead{} & \colhead{[kpc]} & \colhead{[kpc]} & \colhead{} & \colhead{} \\
\colhead{(1)} & \colhead{(2)} & \colhead{(3)} & \colhead{(4)} & \colhead{(5)} &\colhead{(6)} & \colhead{(7)} & \colhead{(8)} & \colhead{(9)} \\ 
}
\startdata
$(1)\, \log\left(R_{e,c} [\mathrm{kpc}]\right)<0.568\times\log\left(M_\ast[M_\sun]\right)$\tablenotemark{c} & \citet{Damjanov2014} & 37 (90\%)\tablenotemark{d} & 10.42 & 0.27 & 1.13 & 0.65 & 0.24 & 0.66\tablenotemark{e}\\
$-5.74$& This work &&&&&&&\\
\hline
$(2)\, \Sigma_{1.5} \equiv log\left(\frac{M_\ast}{R_{e,c}^{1.5}}\left[\frac{M_\sun}{\mathrm{kpc}^{1.5}}\right]\right)>10.3$ & \citet{Barro2013}& 733 (46\%)\tablenotemark{f} & 10.87 & 0.38 & 2.07 & 1.33 & 0.20 & 0.80 \\
& \citet{Poggianti2013a} &&&&&&&\\
\hline
(3)\, $\log\left(R_{e,c}\left[\mathrm{kpc}\right]\right)<0.54\times\log\left(M_\ast\left[M_\sun\right]\right)$ & \citet{Cassata2011} & 840 (52\%)\tablenotemark{g} & 10.83 & 0.37 & 1.83 & 1.02 & 0.20 & 0.80 \\
$-5.5$& \citet{Cassata2013} &&&&&&&\\
\hline
$(4)\, \log\left(R_{e,c}\left[\mathrm{kpc}\right]\right)<0.54\times\log\left(M_\ast\left[M_\sun\right]\right)$ & \citet{Cassata2011} & 82 (5\%)\tablenotemark{g} & 10.82 & 0.38 & 1.04 & 0.58 & 0.22 & 0.80 \\
$-5.8$& \citet{Cassata2013} &&&&&&&\\
\hline
$(5)\, \frac{R_e}{\left(M_\ast\left[10^{11}M_\sun\right]\right)^{0.75}}<2.5$ & \citet{vanderWel2014} & 215 (13\%)\tablenotemark{f}& 11.09 & 0.34 & 2.79 & 1.61 & 0.22 & 0.80 \\
\hline
$(6)\, \frac{R_e}{\left(M_\ast\left[10^{11}M_\sun\right]\right)^{0.75}}<1.5$ & \citet{vanderWel2014}  & 19 (1\%)\tablenotemark{f}& 11.38 & 0.27 & 2.72 & 1.08 & 0.22 & 0.76 \\
\enddata
\tablenotetext{a}{Size of the sample is given both as the total number of compact systems and as the fraction of a parent sample.}
\tablenotetext{b}{Effective radius is circularized in all definitions but the ones used in van der Wel et al. 2014 (Definitions 5 and 6).}
\tablenotetext{c}{This definition is based on the best fit linear relation between the size and the mass of $z\sim0$ quiescent galaxies in our SDSS comparison sample. Compact galaxies are defined to be at least a factor of two smaller than SDSS galaxies of the same mass.}
\tablenotetext{d}{The total number of galaxies in the parent sample - intermediate-redshift quiescent COSMOS galaxies that are classified as point sources in the SDSS photometric database - is 41 (Section~\ref{sizes}).}
\tablenotetext{e}{The cutoff magnitude for this sample of $r(AB)=21.8$~mag strongly affects the upper redshift limit of the selected sample.}
\tablenotetext{f}{The total number of galaxies in the parent sample - intermediate-redshift quiescent COSMOS galaxies with spectroscopically confirmed redshifts and $I(AB)<23$~mag - is 1599 (Section~\ref{COSMOSsample}).}
\tablenotetext{g}{The total number of galaxies in the parent sample - intermediate-redshift quiescent COSMOS galaxies with spectroscopically confirmed redshifts and $I(AB)<23$~mag scaled to Salpeter IMF and  updated version of \citet{Bruzual2003} models - is 1613.}
\end{deluxetable*}
 
The fraction of compact systems in each parent sample depends on the adopted definition. In \citet{Damjanov2014} we pre-select compact galaxy candidates by looking for $r(AB)<21.8$~mag systems that are classified as point sources in the SDSS photometric database, but that show spectroscopic signatures of redshifted quiescent galaxies. All of the candidates with available high-resolution imaging are at least two times smaller than a typical quiescent SDSS galaxy of equivalent mass at $z\sim0$. If we use the same criteria to select compact galaxies in the quiescent sample of the COSMOS field (Section~\ref{sizes}), almost all of candidates are indeed more than two times smaller than their massive SDSS counterparts at $z\sim0$ (Definition 1 in Table~\ref{tab1}). Interestingly, this sample contains mostly lower-mass systems, with an average stellar mass of $\langle M_\ast\rangle\sim2.6\times10^{10}\, M_\sun$.  Because of the stellar mass -- size relation, massive galaxies can be compact even if they appear extended in poor seeing conditions (although their size measurement based on such imaging may not be accurate). Thus we extract different samples using all of the intermediate-redshift quiescent galaxies in the COSMOS field with available spectroscopic redshifts, stellar masses, and structural parameters. 

The upper limit on the mass-normalized galaxy size is crucially important in determining the number of galaxies in each sample. Cutoffs that are close to the size-mass relation defined by the local sample produce subsets that contain half of the total quiescent galaxy sample (Definitions 2 and 3 in Table~\ref{tab1}, \citealt{Barro2013} and compact galaxies in \citealt{Cassata2011}). In contrast,  defining ultra-compact galaxies according to \citet{Cassata2011} (Definition 4 in Table~\ref{tab1}) puts these systems 0.4~dex (or lower) below similarly massive local systems. Although this approach naturally produces a much smaller compact sample, mass distributions of systems classified according to  Definitions 2, 3, and 4 are remarkably similar because all three cutoffs are based on the linear log(size)-log(stellar mass) relations with similar slopes. 

The steepest $d(\log R_e)/d(\log M_\ast)$ gradient \citep[Definitions 5 and 6 in Table~\ref{tab1}, ][]{vanderWel2014} produces a compact sample shifted towards higher stellar masses and larger sizes. In selecting compact galaxy samples \citet{vanderWel2014} use galaxy size measured along the galaxy major axis; they do not circularize it. This approach is the  major factor accounting for the discrepancy between the linear coefficient of Definitions 5 and 6 and others listed in Table~\ref{tab1}. Although mass distributions (and subsequently size distributions) vary among compact samples listed in Table~\ref{tab1}, all samples with the same limiting magnitude ($I(AB)_{lim}=23$~mag) cover the full range of redshifts that we explore. None of the tested definitions produces a sample of compact systems which disappear at low redshift. 
   
\section{The abundance of compact massive galaxies in COSMOS}\label{abundanceexample}

In order to robustly estimate the number density of compact galaxies at intermediate redshift we need: a) a large enough survey area to provide numerous objects and to minimize the effect of cosmic variance,  b) high-quality imaging to enable reliable size measurements of these extreme systems, and c) spectroscopy for secure redshift estimates. The COSMOS field, with HST ACS imaging available for 1.65~deg$^2$ and its (sparse) spectroscopic coverage, is currently one of the best regions for this type of analysis. One of compact samples (Section~\ref{compactsamples}) serves here to illustrate all of the steps in calculating compact galaxy number densities. We demonstrate that the trend in number density with redshift is consistent with an approximately constant abundance in the redshift range $0.2<z<0.8$.        

\subsection{Properties of the compact sample}\label{compprop}

\begin{figure}
\begin{centering}
\includegraphics[scale=0.25]{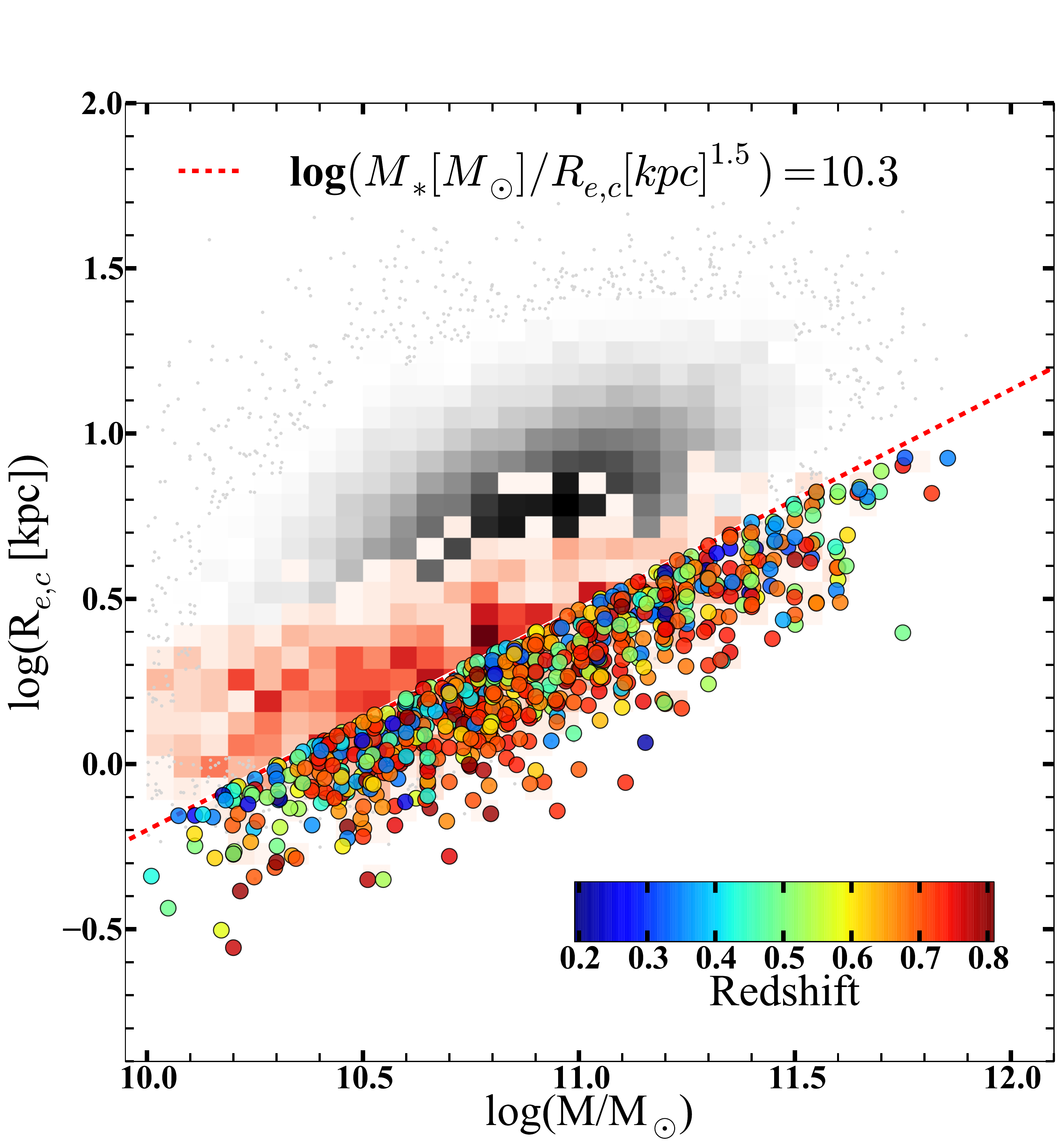}
\caption{Circularized effective radius of single-profile S\'ersic models as a function of stellar mass for the COSMOS compact galaxies (circles color-coded by redshift). The red 2D histogram represents the distribution of the parent quiescent sample at intermediate redshifts and the red dashed line shows the compactness cutoff.  The gray 2D histogram shows a set of $z\sim0$ SDSS massive quiescent galaxies. \label{f3}}
\end{centering}
\end{figure}

\begin{figure}
\begin{centering}
\includegraphics[scale=0.25]{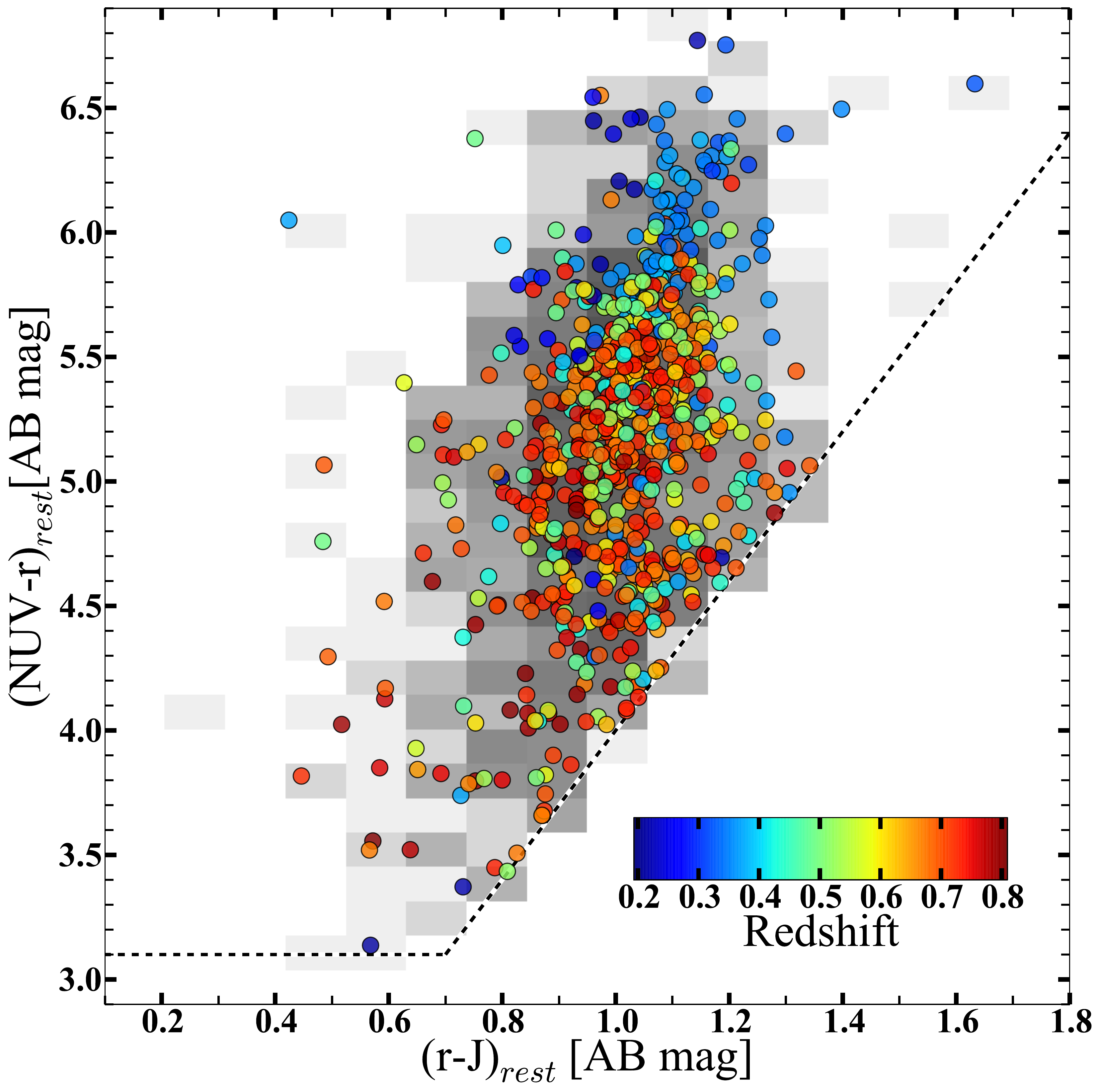}
\caption{Rest-frame $NUV-r$ vs. rest-frame $r-J$ color  for the COSMOS sample of compact galaxies (circles color-coded by redshift). The gray 2D histogram show the underlying distribution of the parent sample. The black dashed lines show the boundary between star-forming and quiescent systems \citep{Ilbert2010}. \label{f4}}
\end{centering}
\end{figure}

\begin{figure*}
\begin{centering}
\includegraphics[scale=0.325]{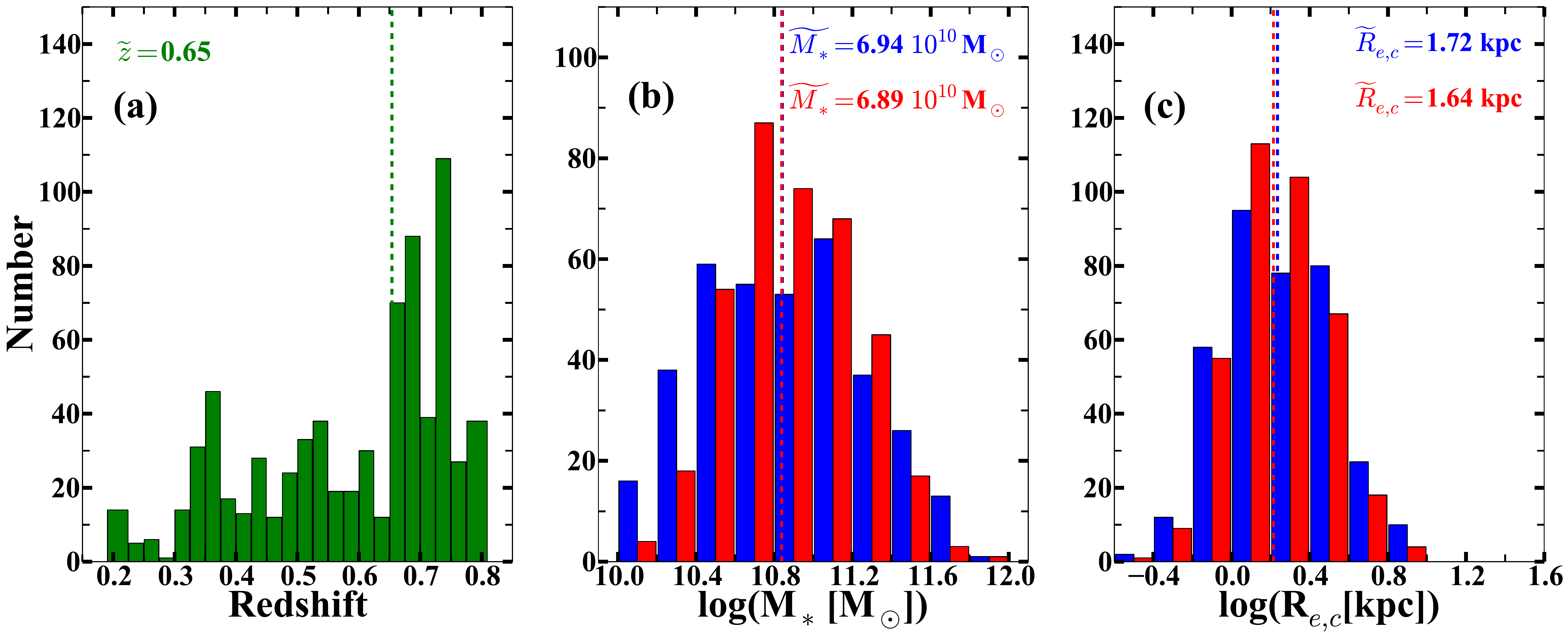}
\caption{ Distribution of spectroscopic redshifts (a), stellar masses (b) and circularized effective radii (c) of the intermediate redshift massive quiescent galaxies in COSMOS classified as compact based on their pseudo-stellar mass surface density $\Sigma_{1.5}$ (Definition~2 in Table~\ref{tab1}). Dashed lines show median values for each quantity. We show stellar mass and effective radius panels subsamples below (blue) and above (red) median redshift ($\widetilde z =0.65$).\label{f5}}
\end{centering}
\end{figure*}

We use Definition~2 from Table~\ref{tab1} to extract a sample of compact massive quiescent galaxies based on their pseudo-stellar mass surface density $\Sigma_{1.5}$. This criterion was introduced by \citet{Barro2013} who applied it to estimate the number density evolution of compact galaxies at $0.5<z<3$ in the CANDELS survey \citep{Grogin2011,Koekemoer2011}. \citet{Poggianti2013a} employed the same definition to show that the number density of compact massive systems in the nearby Universe ($z<0.1$ galaxies drawn from the PM2GC survey) is very similar to their abundance at high redshift. 

Figure~\ref{f3} shows the position of selected compact systems (circles colored by redshift) in the size-stellar mass parameter space compared with the distribution of all massive quiescent  COSMOS systems at $0.2.<z<0.8$ (red two-dimensional (2D) histogram). For the comparison $z\sim0$ sample (gray scale 2D histogram) we employ a set of spectroscopically confirmed quiescent galaxies from SDSSS DR7 with available size measurements \citep{Damjanov2014, Simard2011}. We derive stellar masses for the $z\sim0$ quiescent sample using the same spectral energy distribution fitting method as \citet{Ilbert2013}. The COSMOS compact sample contains 46\% of the parent quiescent sample at intermediate redshift (Table~\ref{tab1}). These dense intermediate-redshift systems are smaller than $95\%$ of local massive passive galaxies.

In rest-frame color-color space the compact sample coincides, by design, with the locus of quiescent systems (Figure~\ref{f4}, Section~\ref{COSMOSsample}). Compact galaxies at lower redshifts show redder $r-J$ and $NUV-r$ rest-frame colors because they are dominated by older stellar populations. Both galaxy colors become bluer (i.e., galaxies are younger) as their redshift increases (see also Paper II). Although this trend is clear (Figure~\ref{f4}), there is also some evidence of mixing. A small fraction of compact galaxies at the lower redshift end - $\sim5\%$ of all compacts at $z\leqslant0.4$ -  show bluer colors (with $NUV-r<4.5$ and $r-J<1$) and appear to be more recently quenched. On the other hand, $\sim12\%$ of compact systems  at $z\geqslant0.6$ have very red rest-frame colors - $NUV-r>5.5$ and $r-J>1$. 

Figure~\ref{f5} shows the distribution of redshifts, masses and sizes for the compact sample. The median redshift of the compact galaxies ($\widetilde z_\mathrm{compact}=0.65$, Figure~\ref{f5}a) exceeds that for the parent sample ($\widetilde z_\mathrm{parent}=0.53$, Figure~\ref{f1}a). If compact systems are preferentially located in overdense regions, this difference may be related to the presence of the largest dense structure in the COSMOS field at $z\sim0.7$ We can test this conjecture only after correcting both the parent and compact samples for spectroscopic incompleteness.   

The size distributions of the compact sample below and above its median redshift are very similar. The median compact sizes in two redshift bins are almost identical ($\widetilde R_{e,c}\sim1.7$~kpc, Figure~\ref{f5}c); they are naturally smaller than the sizes for the parent sample (Figure~\ref{f1}c). 

\subsection{Observational selection effects}\label{obseff}

\begin{figure*}
\begin{centering}
\includegraphics[scale=0.3]{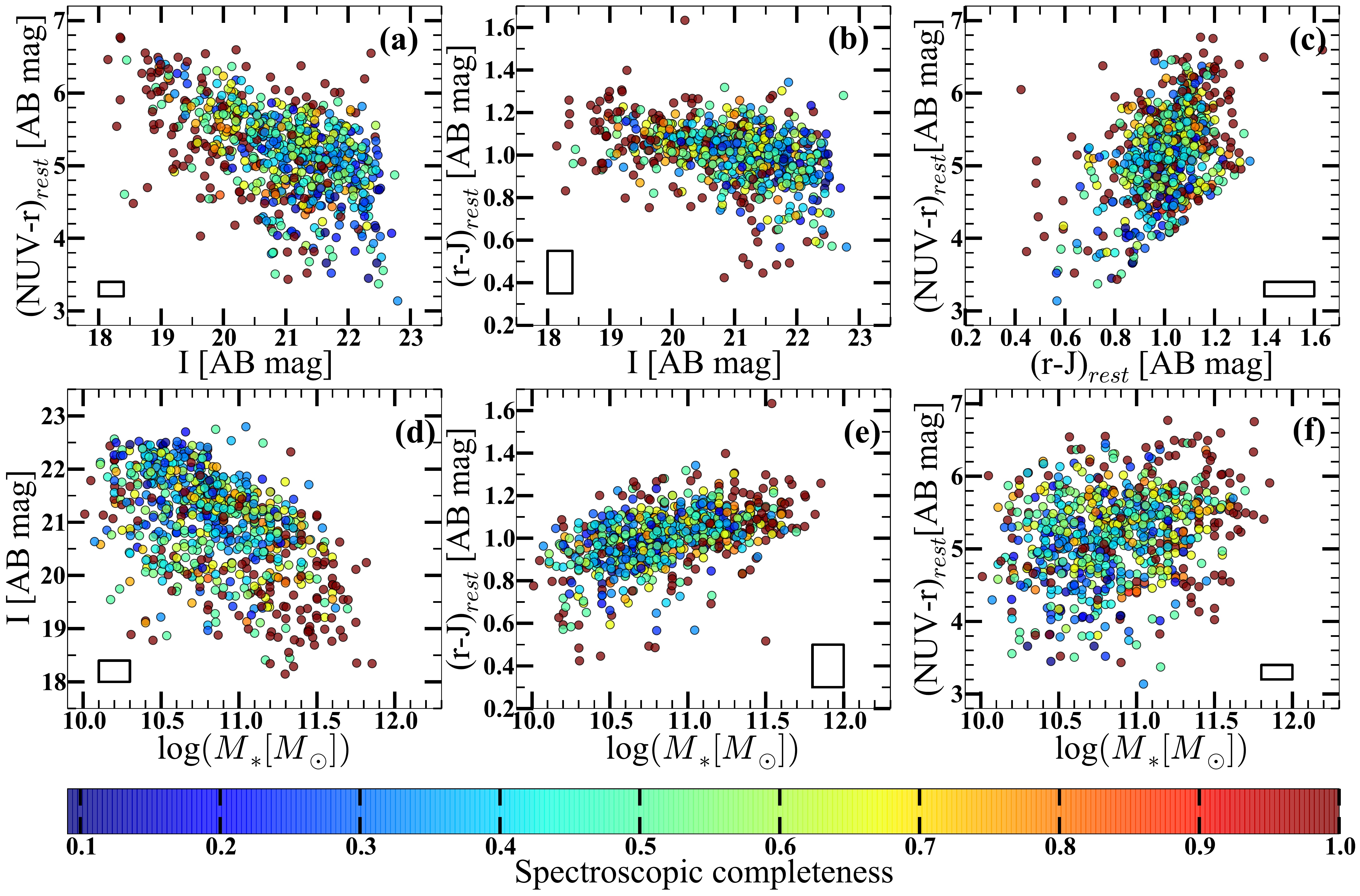}
\caption{Positions of compact galaxies (selected using Definition~2 from Table~\ref{tab1}) in six projections of the parameter space we use to calculate spectroscopic completeness factors. These parameters include: rest-frame $NUV-r$, observed $I$ magnitude, rest-frame $r-J$, and galaxy stellar mass $M_\ast$. Circles color-coded by the spectroscopic completeness factor represent compact systems. The open rectangle in each panel shows the projected size of the cell that we use to calculate the spectroscopic completeness level. See the text for details. \label{f6}}
\end{centering}
\end{figure*}

\begin{figure}
\begin{centering}
\includegraphics[scale=0.2]{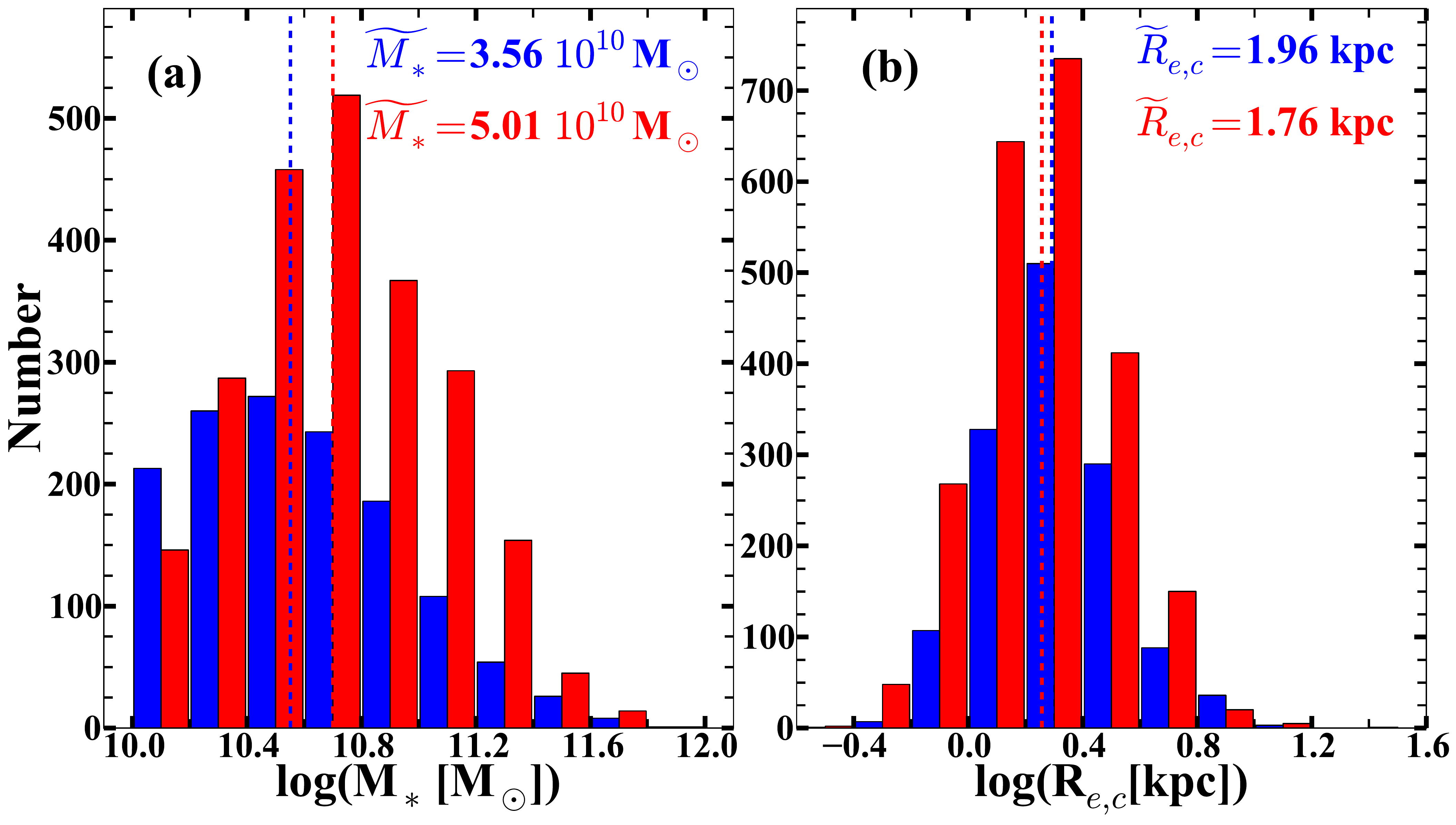}
\caption{Corrected distribution of stellar masses (a) and circularized effective radii (b) for a sample of intermediate redshift massive quiescent galaxies in the COSMOS field at low redshift ($0.2\leqslant z<0.5$; blue) and high redshift ($0.5\leqslant z\leqslant 0.8$; red). Dashed lines indicate median values for each quantity. See the text for details. \label{f7}}
\end{centering}
\end{figure}

\begin{figure}
\begin{centering}
\includegraphics[scale=0.2]{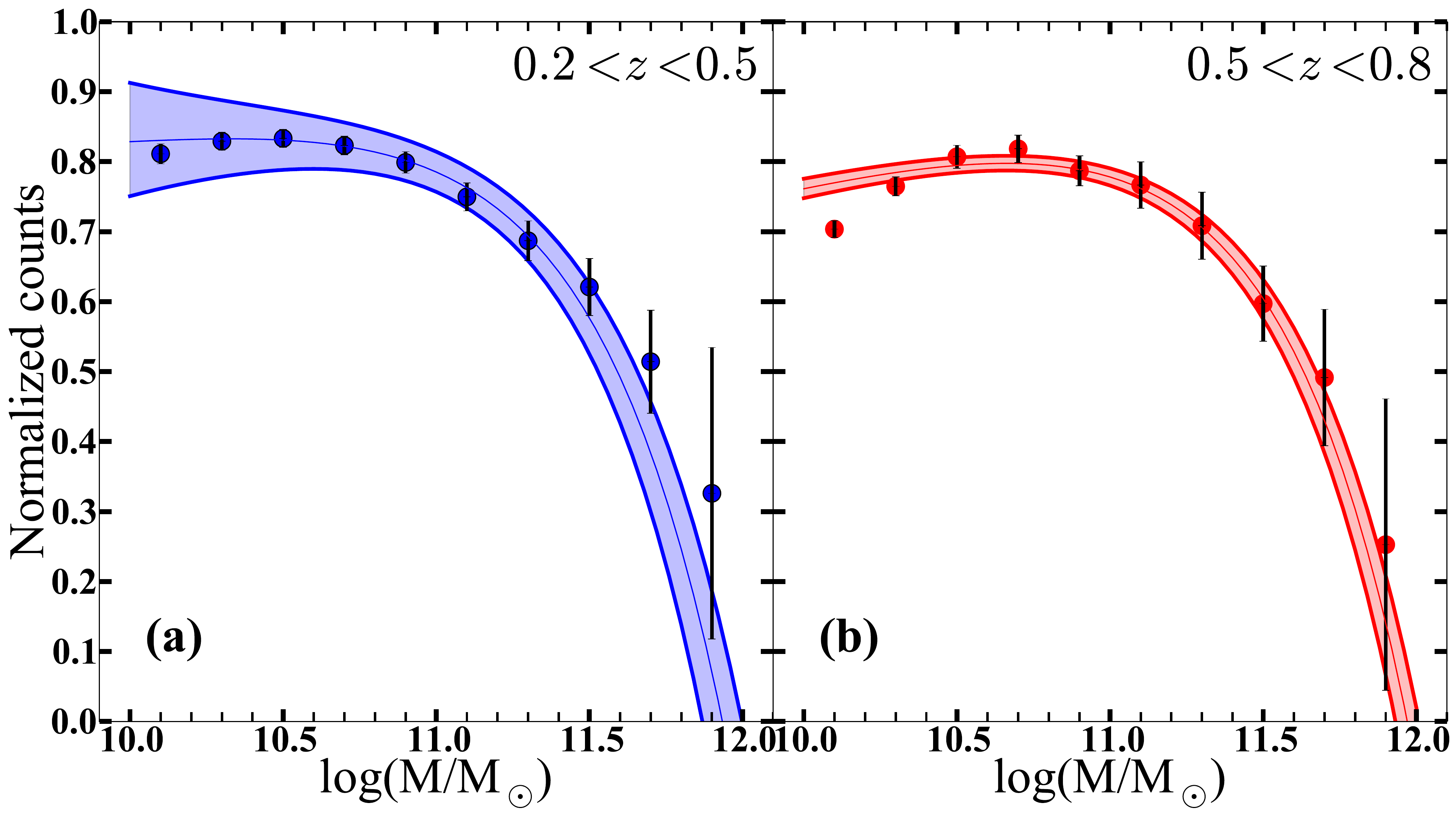}
\caption{ Mass function of the corrected parent quiescent sample based on the COSMOS spectroscopic dataset (points) compared to the mass function defined by the COSMOS quiescent systems with photometric redshifts \citep[][,  shaded regions]{Ilbert2013}. (a) and (b) panels correspond to the low-$z$ and high-$z$ subsamples, respectively. \label{f8}}
\end{centering}
\end{figure}

The sampling by spectroscopic surveys targeting the COSMOS field is sparse at intermediate redshift and we must account for this observational incompleteness to derive compact galaxy number densities. In addition, a magnitude limited sample with available size measurements \citep{Sargent2007} can have a mass distribution that departs from the true mass function of quiescent systems in the lowest mass bins ($M_\ast\sim10^{10}\, M_\odot$) at higher redshifts ($z>0.5$). Here we describe the two corrections required to calculate the intermediate-redshift number densities of massive compact systems. 

To estimate the fraction of compact galaxies with available spectroscopic redshifts in the underlying quiescent population, we use a method based on the position of our spectroscopic targets in color-color-magnitude space \citep{Damjanov2014}. Here we add galaxy mass as the fourth dimension of the parameter space we explore. For each galaxy in the compact sample the (in)completeness correction factor is the ratio between the number of spectro-photometric and the number of photometric extended sources (i.e., galaxies) with similar colors, apparent magnitude and stellar mass as the selected system. This approach is appropriate if the spectroscopic sample is a compilation of sparse surveys with different selection criteria, such as our sample of COSMOS galaxies with spectroscopic redshifts (Section~\ref{COSMOSsample}).   

We construct the four-dimensional (4D) parameter space using galaxy rest-frame colors $NUV-r$ and $r-J$, that are also used to discriminate between star-forming and quiescent galaxies (Section~\ref{COSMOSsample}), apparent $I-$band magnitude, and stellar mass. For each compact galaxy we define a cell with $\Delta(NUV-r)=\Delta(r-J)=0.2$~mag, $\Delta I=0.4$~mag, $\Delta \log(M_\ast [M_\odot])=0.2$ sides centered at the position of the compact target in this parameter space. We count galaxies with measured sizes located within a given cell and divide their number by the number of galaxies with additional spectroscopic information. This ratio is then the estimated spectroscopic completeness correction at the position of selected compact target. 

We illustrate our method in the panels of Figure~\ref{f6} that show compact objects (circles) in the projections of the ($NUV-r$ rest-frame color - $r-J$ rest-frame color - $I$ apparent magnitude - stellar mass) parameter space. Each circle is color-coded by the spectroscopic correction we apply to the corresponding object. The open rectangles depict the size of the 4D cell in all six projections. In practice we construct a separate 4D cell around the position of each compact target defined by its stellar mass, rest-frame colors, and $I-$band magnitude.

The size of the cells we use to estimate the correction for spectroscopic completeness is set by the number and mass distribution of galaxies in the photometric sample. Using a grid of cell sizes, we apply a range of corrections for spectroscopic incompleteness to the parent sample of quiescent galaxies with measured sizes. As a result, the same galaxy has different incompleteness factors associated with it in different corrected parent samples.  We divide each corrected parent sample in two subsets defined by spectroscopic redshift ($0.2\leqslant z<0.5$ and $0.5\leqslant z\leqslant0.8$). For both redshift intervals we require the corrected number of galaxies in each narrow mass bin ($\Delta \log(M_\ast [M_\odot])=0.2$) to be close to (but not to exceed) the number of similarly massive galaxies with photometric redshifts in the same broad redshift range. This procedure selects the optimal cell size which we then use to derive correction factor for each system in our compact COSMOS sample.  The resulting spectroscopic completeness factors cover the range $0.1-1$ (as shown in Figure~\ref{f6}) with an average value (and standard deviation) of $0.57\pm0.26$. 

In addition to its spectroscopic incompleteness, our compact sample may be incomplete at the lower mass end because of the magnitude limit of the catalog containing the galaxy structural parameters ($I_{lim}(AB)=23$). We account for this additional incompleteness by dividing galaxies from the  COSMOS photometric catalog \citep{Ilbert2013}  into low-redshift ($0.2\leqslant z_{phot}<0.5$) and high-redshift ($0.5\leqslant z_{phot}\leqslant0.8$) subsets and calculating the fraction of galaxies with $I(AB)<23$ in the underlying quiescent galaxy population in narrow mass bins ($\Delta \log(M_\ast [M_\odot])=0.2$) for both subsets. This ratio is less than unity only in two mass bins of the high-$z$ subsample: it is 0.68 at $M_\ast\sim1.2\times10^{10}\, M_\odot$ and 0.89 at $M_\ast\sim2\times10^{10}\, M_\odot$. To test the method we apply the correction to the parent galaxy sample, which we first correct for the spectroscopic incompleteness. 

Figure~\ref{f7} shows the mass and size distributions of the parent sample of quiescent COSMOS galaxies with spectroscopically confirmed redshifts, corrected for the incompleteness due both to spectroscopic sampling and magnitude limit of the morphological catalog. Although the shape of the size distribution (Figure~\ref{f7}b) is the same as for the input sample (shown in Figure~\ref{f1}), the galaxy mass distribution of the corrected sample is flatter in the low-mass regime ($10^{10}\, M_\odot<M_\ast<5\times10^{10}\, M_\odot$), at least for systems in the low-$z$ subsample (blue histogram in Figure~\ref{f7}a). 

The number of galaxies of the high-$z$ galaxy subset still declines in the lowest mass bins  (red histogram in Figure~\ref{f7}a). To check whether the corrected mass function of our spectroscopic parent sample corresponds to the independently observed mass function of COSMOS quiescent galaxies spanning the same redshift range \citep{Ilbert2013}, we compare the shapes of two distributions in Figure~\ref{f8}. In the low-$z$ regime our corrected parent spectroscopic sample produces a mass function (blue circles) of the same shape as the mass function of the COSMOS photo-$z$ sample (blue shaded region). In the high-$z$ regime, the mass function of our corrected subsample departs somewhat from the shape defined by the  COSMOS quiescent photo-$z$ sample only in the lowest mass bin ($M_\ast(0.5\leqslant z<\leqslant0.8)\sim10^{10}\, M_\odot$). At all stellar masses $M_\ast\gtrsim2\times10^{10}\, M_\odot$ our spectroscopic and magnitude related completeness factors provide a corrected parent sample of $0.2\leqslant z\leqslant0.8$ quiescent systems with mass functions that agree extremely well with the observed mass functions based on an independent, detailed analysis of the COSMOS quiescent galaxies with photometric redshifts.  

Finally, we test the corrections applied to the spectroscopic parent sample by comparing number densities based on corrected galaxy counts in $\Delta z=0.2$ redshift bins within the $0.2<z<0.8$ redshift interval with the number density evolution of massive quiescent galaxies from the COSMOS photometric sample \citep[][right panel of their Figure~6]{Carollo2013}. We find excellent agreement, with a positive $\sim30\%$ offsets in the redshift bin (centered at $z=0.7$) covering the largest known overdensity in the COSMOS field, for which \citet{Carollo2013} make a correction. This test confirms that the spectroscopic and magnitude related completeness factors do not overestimate the absolute number of COSMOS galaxies in the corrected spectroscopic parent sample of massive quiescent systems in the $0.2<z<0.8$ redshift range.

\subsection{The number density of compact galaxies}\label{res}
 \begin{figure*}
\begin{centering}
\hspace*{-0.35in}
\includegraphics[scale=0.4]{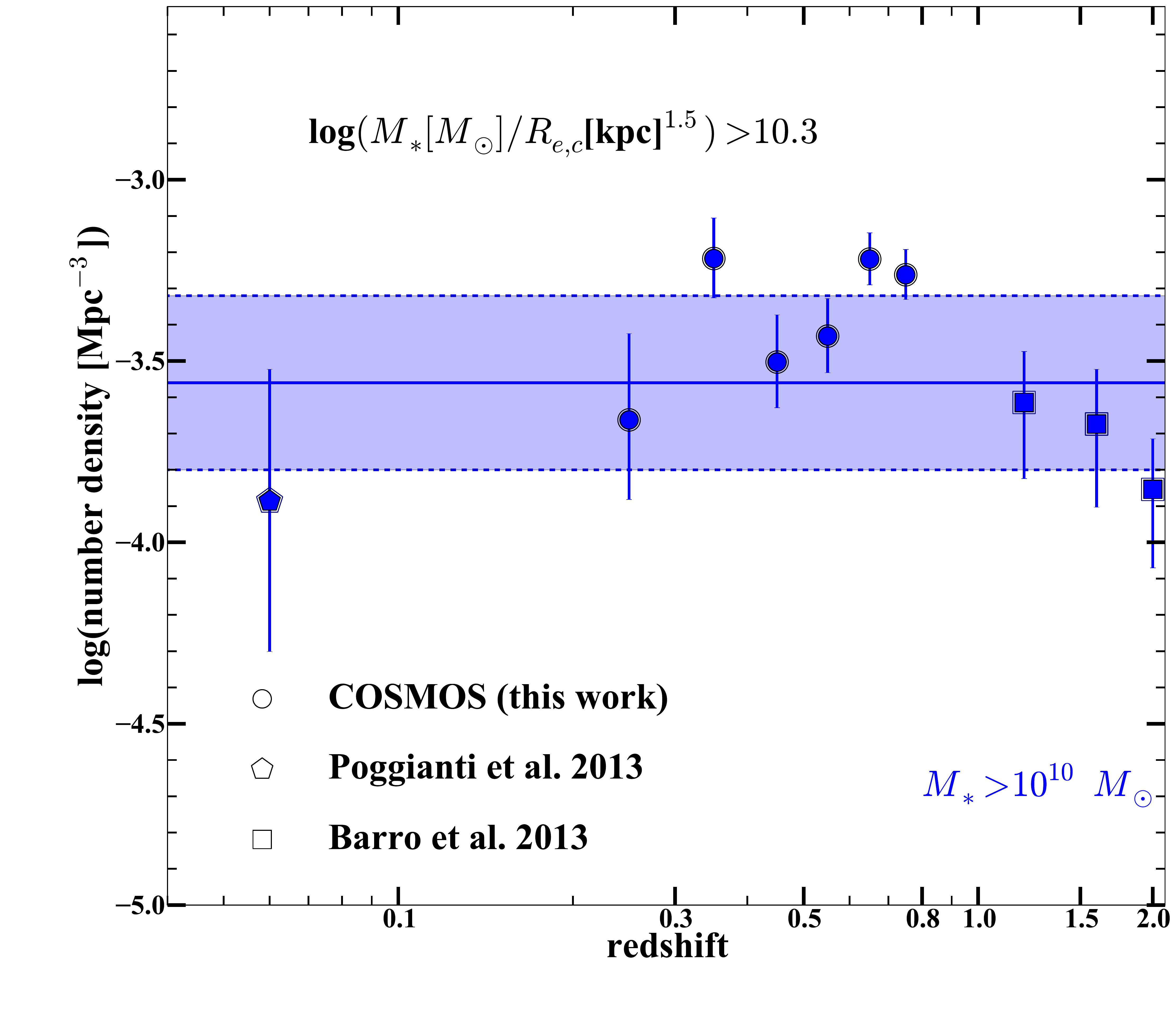}
\caption{Number density of $M_\ast>10^{10}\, M_\odot$ compact galaxies with $\Sigma_{1.5}>10.3\, M_\odot\, \mathrm{kpc}^{-1.5}$ (Definition 2 in Table~\ref{tab1}) as a function of redshift for the COSMOS intermediate-redshift compacts (circles), a set of galaxies at $z\lesssim0.1$ \citep[diamonds, PM2GC sample, ]{Poggianti2013a}, and for high-redshift CANDELS targets \citep[squares,][]{ Barro2013}. The blue line and the blue shaded area correspond to the average number density ($\pm1\sigma$) of compact galaxies at $0<z<2$.\label{f9}}
\end{centering}
\end{figure*}
 
 \begin{figure}
\begin{centering}
\hspace*{-0.35in}
\includegraphics[scale=0.26]{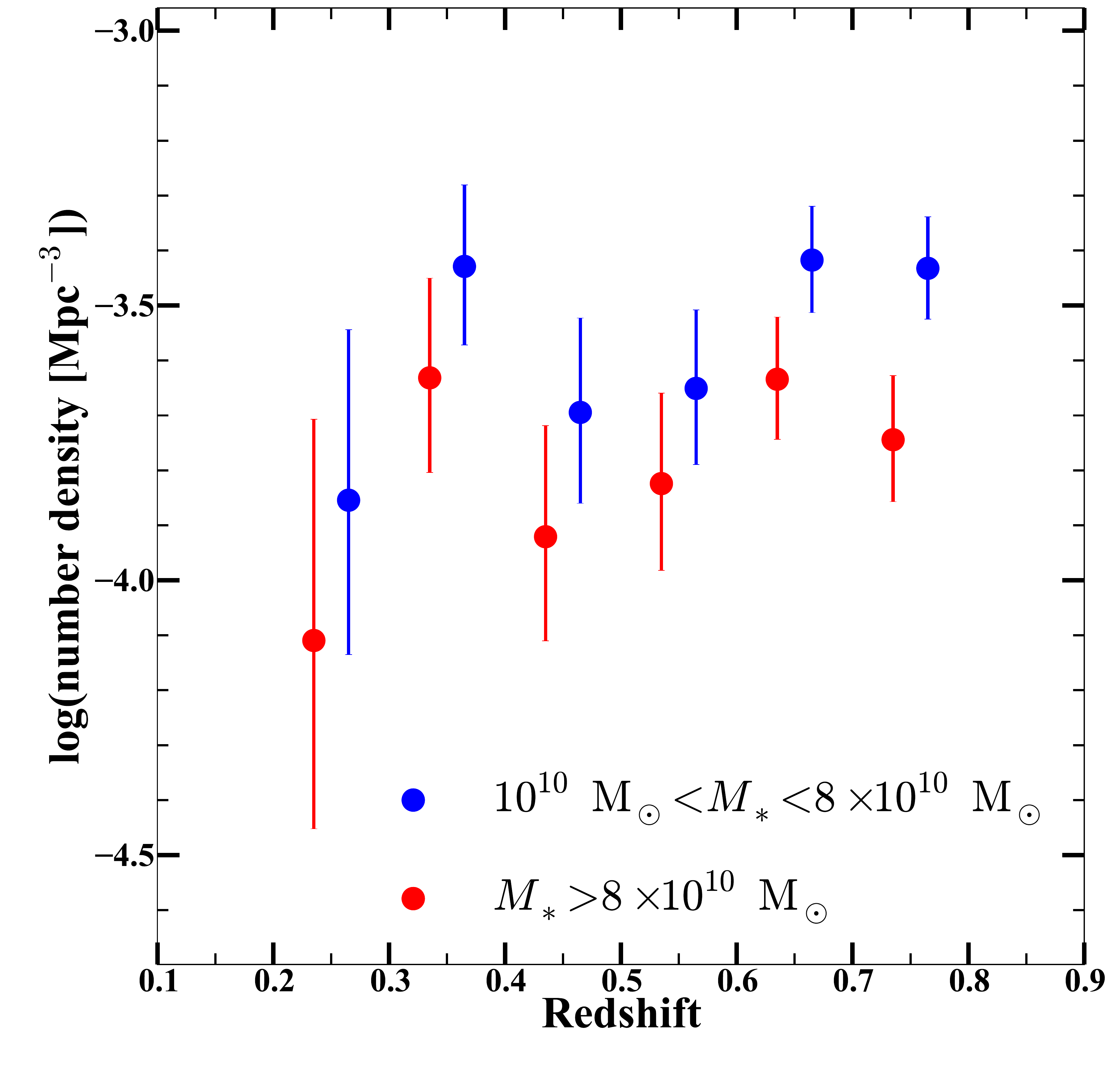}
\caption{ Number density of compact galaxies with $\Sigma_{1.5}>10.3\, M_\odot\, \mathrm{kpc}^{-1.5}$ (Definition 2 in Table~\ref{tab1}) as a function of redshift for the COSMOS sample. The two subsamples are based on galaxy stellar mass. The upper mass limit of the low-mass subset (blue circles) and the lower mass limit of the high-mass subset (red circles) correspond to the characteristic mass of the COSMOS mass function for quiescent galaxies at $0.2<z<0.8$ \citep{Ilbert2013}.  \label{f10}}
\end{centering}
\end{figure}

We apply the correction factors from Section~\ref{obseff} to the spectroscopic subsample of compact intermediate-redshift galaxies in COSMOS selected according to Definition~2 (Table~\ref{tab1}). We then divide the corrected number of compact systems into narrow redshift bins ($\Delta z=0.1$) and compute the corresponding comoving volume element to estimate the abundance of compact galaxies as a function of redshift in the range $0.2\leqslant z\leqslant0.8$. 
 
The uncertainty in the number densities of compact COSMOS galaxies includes errors in the estimated stellar mass, errors in the size measurement, and the confidence limits from Poisson statistics. We build a set of Monte Carlo simulations where we vary both the stellar mass and the size of our selected objects taking into account average ($1\sigma$) errors of $\Delta(\log(M_\ast[M_\odot]))=\Delta(\log(R_{e(,c)}[\mathrm{kpc}]))=0.1$dex. The error in stellar mass is dominated by the uncertainty due to the choice of stellar library \citep{Santini2015}. The error in size corresponds to the maximum difference between the single-S\'ersic effective radius we measure using Galfit and the corresponding size given in the COSMOS morphological catalog \citep[][see Figure~\ref{f2}]{Sargent2007}. The set of 1000 realizations provides a range of number densities for each redshift bin and lower stellar mass limit. We fold standard deviations of the resulting number density distributions into the total uncertainty estimates together with the Poisson confidence limits corresponding to the Gaussian $1\sigma$ errors \citep{Gehrels1986}. The final error budget is dominated by the confidence limits based on Poisson statistics.

We compare the resulting evolution of compact galaxy number density from $z=0.8$ to $z=0.2$ with the values from \citet{Poggianti2013a} at $z\sim0$ and  \citet{Barro2013} at $z>1$ using the same compactness criterion (Figure~\ref{f9}). At the same stellar mass limit ($M_\ast>10^{10}\, M_\odot$, blue symbols in Figure ~\ref{f9}) and with the large uncertainty, the abundance of compact systems increases by 0.2~dex from $z\sim0$ to $z\sim1.5$. The observed number densities are also consistent with a roughly constant value of $\sim2.7^{+2.0}_{-1.2}\times10^{-4}$~Mpc$^{-3}$ in this redshift range (see Section~\ref{cgd} and Table~\ref{tab2}). Thus the number density of compacts in the local universe is very similar to the value reported at high redshift  \citep{Poggianti2013a}. The largest deviations from this constant trend (and the largest absolute values of compact galaxy number densities) occur in the redshift bins with the largest of overdensities in the COSMOS field (at $z\sim0.35, 0.65, 0.75$, see Section~\ref{COSMOSsample}). 

To further probe the number density evolution of the low-mass and high-mass compact systems, we divide the sample into two subsets based on the characteristic stellar mass of the COSMOS mass function for intermediate-redshift quiescent galaxies \citep[$M^\ast=8\times 10^{10}\, M_\odot$,][]{Ilbert2013}. Figure~\ref{f10} shows that the number densities of low-mass and high-mass compact galaxies exhibit almost identical patterns, with absolute number densities of low-mass compacts being slightly exceeding (but within $1\sigma$ uncertainty intervals of) the abundances of high-mass compact systems in all redshift bins. The only redshift range where the two number densities differ more than $1\sigma$ is $0.7<z\leqslant0.8$. In this redshift bin the abundance of compact galaxies with stellar masses above the characteristic mass $M^\ast$ is $\sim2$ times lower than the abundance of similarly compact but less massive systems. Interestingly, this redshift range includes the largest structure known in the COSMOS field \citep{Scoville2007, Scoville2013}. 

Overall, the number density of compact galaxies selected based on their pseudo-stellar mass surface density $\Sigma_{1.5}$ (Definition 2 in Table~\ref{tab1}) is approximately constant for $0.2<z<0.8$. The study of the $ZEST+$-measured size distribution of COSMOS galaxies in this redshift range yields similar results: in the stellar mass range covered by the bulk of  quiescent galaxy population ($10.5<\log(M_\ast/M_\sun)<11$)  the abundance of the densest systems does not evolve \citep{Carollo2013}. The abundances of compact systems in the COSMOS field at these redshifts are similar to the values reported for equivalently defined compact galaxies at both lower and higher redshifts. The variations in the number density values within this redshift range correspond to the redshifts of known COSMOS overdensities. The abundance of compact galaxies is approximately constant over this redshift range regardless of the galaxy mass in the range we explore.

\section{Discussion}\label{dis}

There are several observational effects that can alter the size and properties of the selected compact galaxy sample. These issues include the underlying survey size  and the selection criteria used to define the sample. For a survey like COSMOS,  cosmic variance  may produce large deviations from the typical value in the number of (compact) galaxies in any redshift range. The definition of compactness influences the total number of galaxies in the sample and their mass distribution. Here we investigate these two observational issues and their effect on the resulting number densities of compact systems in COSMOS.

\subsection{Cosmic variance}\label{cv}

\begin{figure}
\begin{centering}
\hspace*{-0.35in}
\includegraphics[scale=0.26]{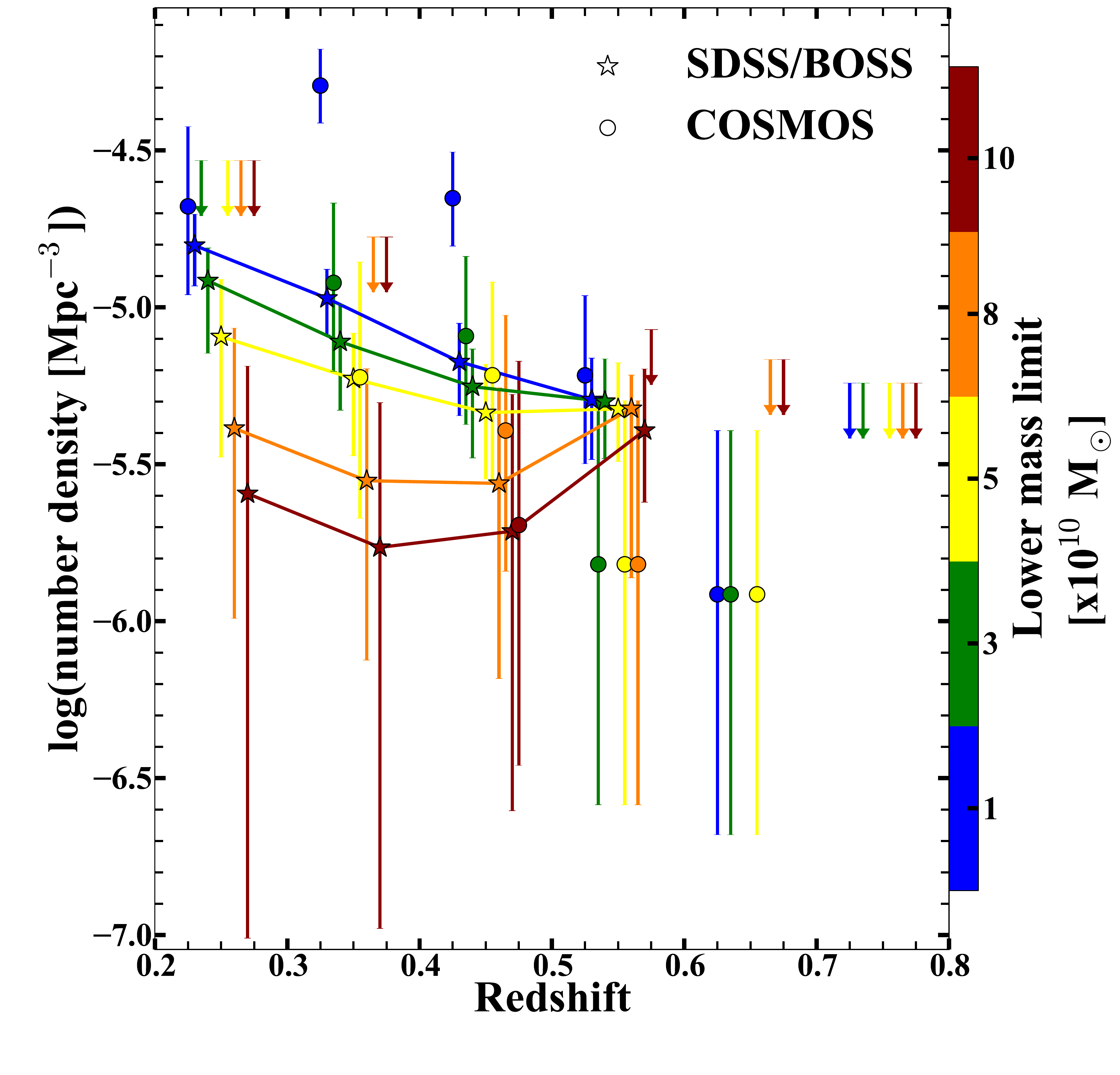}
\caption{ The number density of compact galaxies as a function of redshift for two bright ($r(AB)<21.8$) intermediate-redshift samples: 1) SDSS/BOSS galaxies that appear as point sources in the photometric SDSS database and show signatures of redshifted evolved systems in their BOSS spectra  \citep[][stars]{Damjanov2014}  and 2) COSMOS quiescent galaxies classified as point sources in the photometric SDSS database (circles). Each sample is represented by a set of five subsets color-coded based on the lower mass limit used to select the subset. All upper limits correspond to the Poisson upper limit for the $2\sigma$ confidence level  in the case of non-detection. \label{f11}}
\end{centering}
\end{figure}

Enhanced number densities of compact galaxies in some intermediate-redshift bins correspond to dense structures in COSMOS (Section~\ref{res}). Here we compare lower limits to the number density of compacts determined from a very large volume survey, BOSS, to the similarly defined limits from COSMOS in order to examine the potential impact of cosmic variance on the resulting number density of compacts. 

Using the approach from \citet{Damjanov2014} we select COSMOS sources with $r(AB)<21.8$ that appear as stars in the SDSS images and have spectral energy distributions (SEDs) of intermediate-redshift massive quiescent galaxies. In addition, these objects are extended in the HST ACS images of the COSMOS field and have measured sizes. The final SDSS/COSMOS compact sample contains a subset of these objects that are at least two times smaller than a typical local quiescent galaxy of equivalent stellar mass (Definition~1 from Table~\ref{tab1}). In comparison with other samples listed in Table~\ref{tab1}, the SDSS/COSMOS compact sample has a relatively small number of systems with stellar masses $10^{10}\, M_\odot<M_\ast<10^{11}\, M_\odot$. We use this compact dataset to derive number densities that we label as lower limits in order to distinguish SDSS/COSMOS selection from the complete sample of compact massive galaxies in COSMOS. The number densities of the compact SDSS/COSMOS systems are comparable with measured number densities of intermediate-redshift compact systems based on SDSS/BOSS sample, that are also interpreted as lower limits because of the selection effects associated with the parent BOSS sample \citep{Damjanov2014}.

Figure~\ref{f11} displays the excellent agreement between the two sets of lower limits on compact galaxy number densities. The abundances derived from the SDSS/COSMOS sample (circles) are above and mostly within the uncertainties of the SDSS/BOSS values (stars) for all stellar mass limits (color-coding) in the $0.2<z<0.5$ redshift range. Because the BOSS survey covers an area almost 4000 times larger than COSMOS and gives similar results for a similarly selected sample, we conclude that the effects of cosmic variance are not dominant. Furthermore, this result shows that  the selection algorithm of the BOSS quasar survey \citep{Ross2012}, where we find compact SDSS/BOSS systems, does not exclude any subset of the compact population at these redshifts.  

Interestingly, at some stellar mass limits in this redshift range we do not detect any COSMOS compact galaxies classified as stars in the SDSS imaging (upper limits corresponding to the $2\sigma$ confidence level for non-detection in Figure~\ref{f11}). Intermediate-redshift massive galaxies that are compact enough to be misclassified as point sources in low-resolution images (FWHM(SDSS)$\sim1\farcs5$) are very rare. The search for such extreme objects requires a large volume, like the BOSS survey. 

At the high end of the redshift range covered by the SDSS/COSMOS sample ($0.6<z<0.8$) galaxies become too faint for our SDSS apparent magnitude cutoff. Upper values on the number densities of SDSS/COSMOS compact systems in this redshift regime reflect the limitation of the bright magnitude-limited sample. 

The agreement between the lower limits on the number densities based on SDSS/COSMOS and SDSS/BOSS samples shows that cosmic variance does not dominate the resulting overall trend in the COSMOS compact galaxy number density with redshift.  It also demonstrates that the broad color selection of the BOSS quasar survey is not biased against compact massive intermediate redshift galaxies. This comparison is a reminder that in the intermediate-redshift regime, massive systems may appear extended even in low-quality imaging surveys and still be defined as compact. In order to select all compact quiescent galaxies using a criterion based on the size-stellar mass relation, the parent sample has to include all massive systems with available size measurements. 

\subsection{Compact galaxy definition}\label{cgd}

\begin{figure*}
\begin{centering}
\hspace*{-0.35in}
\includegraphics[scale=0.27]{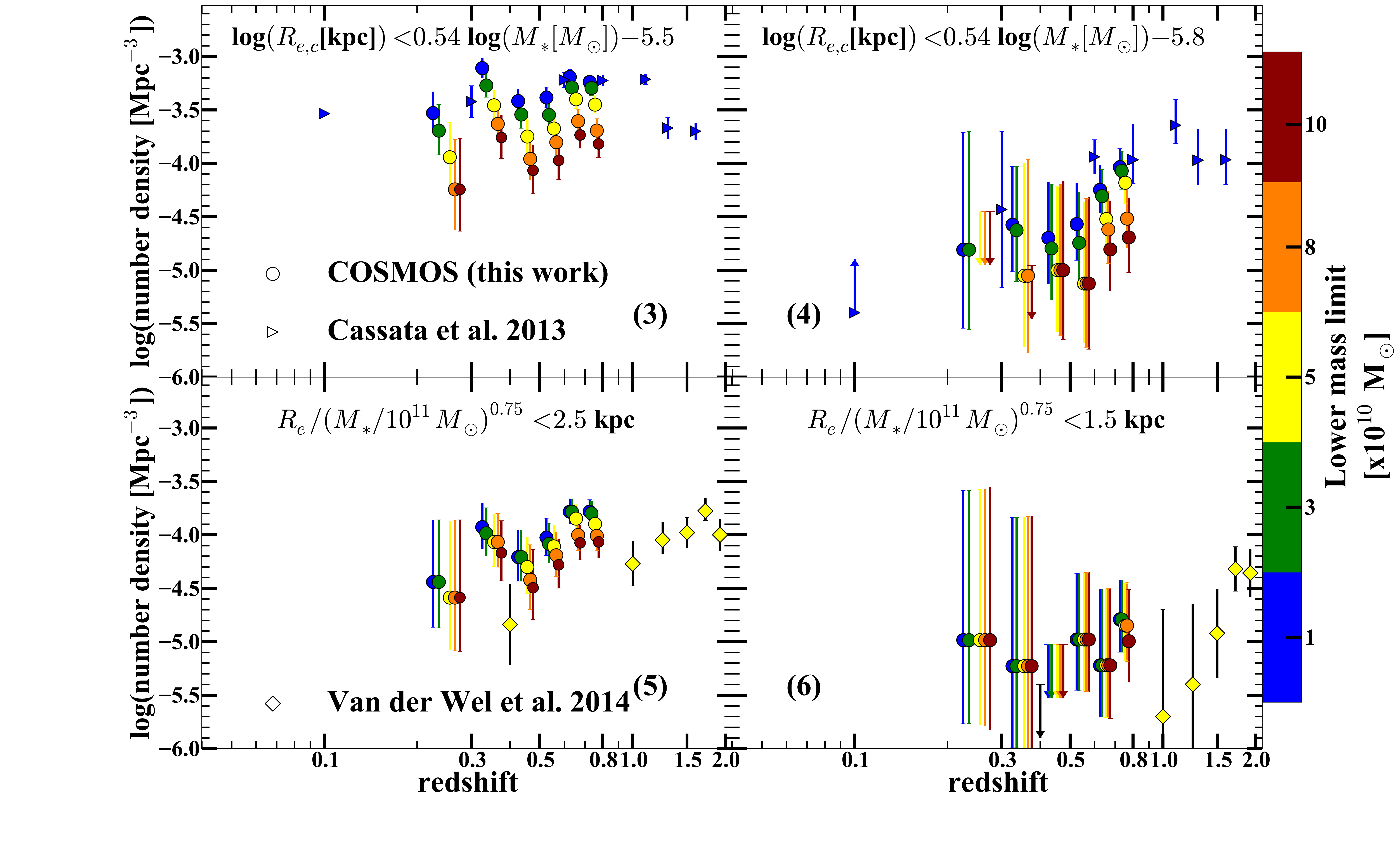}
\caption{Number density of compact galaxies as a function of redshift for different compactness thresholds (numbered as in Table~\ref{tab1}). Each panel contains compact samples defined using the expression shown in that panel. Our results (circles) are clearly similar to the results of other surveys that define compact massive galaxies in the same manner (triangles in the upper panels \citep{Cassata2013}  and diamonds in the lower panels \citep{vanderWel2014}), with the exception of the lowest redshift point of \citet{vanderWel2014} in panel (5). All results are color-coded by the lower limit on the stellar mass of galaxies in the sample.\label{f12}}
\end{centering}
\end{figure*}

\begin{deluxetable*}{lcll}
\tabletypesize{\scriptsize}
\setlength{\tabcolsep}{0.1pt} 
\tablecaption{The evolution of compact galaxy number density \label{tab2}}
\tablewidth{0pt}
\tablehead{ \colhead{Definition\tablenotemark{a}} & \colhead{$M_\ast$} & \colhead{$0.2\leqslant z\leqslant 0.8$\tablenotemark{b,d} }& \colhead{$0<z\leqslant1.5$\tablenotemark{c,d}} \\
\colhead{} & \colhead{$[10^{10}\, M_\sun]$} & \colhead{} & \colhead{} \\
\colhead{(1)} & \colhead{(2)} & \colhead{(3)} & \colhead{(4)} \\ 
}
\startdata
$(2)\, \Sigma_{1.5} \equiv log\left(\frac{M_\ast}{R_{e,c}^{1.5}}\left[\frac{M_\sun}{\mathrm{kpc}^{1.5}}\right]\right)>10.3$ & $>1 $ & $n(z)=10^{(-3.55\pm0.26)}(1+z)^{(1.25\pm1.23)}$ & $n(z)=10^{(-3.43\pm0.23)}(1+z)^{(0.57\pm1.07)}$ \\
$(3)\, \log\left(R_{e,c}\left[\mathrm{kpc}\right]\right)<0.54\times\log\left(M_\ast\left[M_\sun\right]\right)$ &  $>1$ & $n(z)=10^{(-3.34\pm0.26)}(1+z)^{(0.45\pm1.25)}$ & $n(z)=10^{(-3.20\pm0.16)}(1+z)^{(-0.27\pm0.63)}$ \\
 $(4)\, \log\left(R_{e,c}\left[\mathrm{kpc}\right]\right)<0.54\times\log\left(M_\ast\left[M_\sun\right]\right)$ & $>1$ & $n(z)=10^{(-5.61\pm0.27)}(1+z)^{(6.36\pm1.22)}$ & $n(z)=10^{(-4.72\pm0.28)}(1+z)^{(2.82\pm1.12)}$\\
$(5)\, \frac{R_e}{\left(M_\ast\left[10^{11}M_\sun\right]\right)^{0.75}}<2.5$ & $>5$ & $n(z)=10^{(-4.62\pm0.29)}(1+z)^{(3.16\pm1.55)}$ & $n(z)=10^{(-4.02\pm0.19)}(1+z)^{(0.07\pm0.67)}$\\
$(6)\, \frac{R_e}{\left(M_\ast\left[10^{11}M_\sun\right]\right)^{0.75}}<1.5$ & $>5$ & $n(z)=10^{(-5.46\pm0.42)}(1+z)^{(2.26\pm1.90)}$ & $n(z)=10^{(-5.04\pm0.30)}(1+z)^{(0.04\pm1.02)}$\\
\enddata
\tablenotetext{a}{Equivalent to definitions in Table~\ref{tab1}}
\tablenotetext{b}{Based on COSMOS data only.}
\tablenotetext{c}{Based on the COSMOS field number densities combined with equivalently derived number densities from the literature.}
\tablenotetext{d}{The best-fit parameters and their $1\sigma$ uncertainties are formal results of the weighted least square linear regression: $\log(n)=\alpha+\beta\log(1+z)$}
\end{deluxetable*}

Compact galaxy samples are defined based on the galaxy position in the size-stellar mass parameter space (Section~\ref{compactsamples}). The two major factors affecting the size and properties of compact systems are: 1) the slope of the best-fit linear relation between galaxy size and stellar mass at $z\sim0$ (linear coefficient of the selection function), and 2) the distance between the threshold for compactness and the $z\sim0$ size-mass relation (zero-point of the selection function). Although the slope of the local size-mass relation for quiescent systems is well known \citep[$\sim0.6$; e.g.,][]{Shen2003}, even a small change in value may produce a very different mass distribution for the compact quiescent sample. The position of the compactness threshold crucially affects the number of galaxies in the selected sample. We investigate the effects of these two factors on the redshift evolution of compact galaxy number densities for the COSMOS samples (Table~\ref{tab1}).

The absolute value of the compact galaxy abundance depends on the number of objects in the sample (i.e, on the zero point of the selection function). We perform the procedure described in Section~\ref{abundanceexample} using four different criteria for identifying compact galaxies in the COSMOS sample (Definitions 3--6 in Table~\ref{tab1}). Compact COSMOS galaxy subsets corresponding to a more extreme cutoff for compactness (circles in panels (4) and (6) of Figure~\ref{f12}) produce an order of magnitude lower number densities than samples based on more inclusive selection criteria for the same size-mass relation slope (circles in panels (3) and (5) of Figure~\ref{f12}, respectively).

The linear coefficient of the selection function produces the spread between number densities of subsets with different lower mass limits (color-coded circles in figure~\ref{f12}). A shallower slope of the size-mass relation gives a range of abundances that are progressively lower with increasing minimum stellar mass of galaxies in the sample. The number density of compact systems with $M_\ast>10^{10}\, M_\odot$ is a factor of four (0.6~dex) larger than the number density of the most massive compact galaxies with $M_\ast>10^{11}\, M_\odot$ (color-coded circles in Figure~\ref{f9} and in panels (3) and (4) of Figure~\ref{f12}). A selection function with a lower linear coefficient is more likely to include the maximum fraction of objects at the lower end of the parent mass distribution.

A steeper $d(\log R_e)/d(\log M_\ast)$ gradient of the selection function excludes low-mass systems and produces a compact sample shifted to higher stellar mass (Section~\ref{compactsamples}). The resulting number densities for different limiting stellar masses thus exhibit much smaller scatter as a function of the difference between the complete compact sample ($M_\ast>10^{10}\, M_\odot$) and the most massive compact systems, in the range of $0-0.3$~dex (panels (5) and (6) of Figure~\ref{f12}). This difference becomes more prominent in the high-redshift bins that probe lager comoving volumes; these samples are more likely to include rare extremely compact low-mass systems (see panel (6) of Figure~\ref{f12} for the example of selection function with high linear coefficient and low zero-point).

We compare derived number densities of these differently selected samples of compact galaxies in COSMOS with the results of other studies that employ equivalent selection criteria. Definitions 3 and 4 were introduced by \citet{Cassata2011} and we add their updated results for GOODS-South field \citep{Cassata2013} to panels (3) and (4) of Figure~\ref{f12} (blue triangles). For both definitions we find very good agreement between the abundances of compact systems in these two fields. In panels (5) and (6) of Figure~\ref{f12} we compare COSMOS results with measurements based on the sample of equivalently defined compact galaxies in the CANDELS fields \citep[][yellow diamonds]{vanderWel2014}. Here we find agreement for the abundances of the most extreme (and thus rare) compacts (panel (6)) that also have the highest associated uncertainties. For the more inclusive compactness threshold (panel (5)) \citet{vanderWel2014} suggest that the number density of compact systems in CANDELS declines drastically between $z=1$ and $z=0$. As we discuss below, the range of possible number density trends with redshift for the COSMOS $0.2<z<0.8$ dataset in panel (5) of Figure~\ref{f12} is sufficiently large to include the lowest rates of evolution in compact galaxy number density for five CANDELS fields reported by \citet{vanderWel2014}. However, in combination with the $z\gtrsim1$ CANDELS data points, the COSMOS field data suggest more moderate change in the number density of compacts in the range $0.2<z<1.5$.

To test wether the observed number density evolution is consistent with constant abundance of compact galaxies in the intermediate redshift range, we derive power-law relations between number density and redshift for every compact sample in Figures~\ref{f9} and~\ref{f12}. Taking the uncertainties in number densities into account, we perform weighted least square fits 1) to the COSMOS data in the $0.2\leqslant z\leqslant0.8$ range alone, and 2) to combined data points from COSMOS and other fields covering redshift range $0<z\leqslant1.5$. The formal best-fit parameters with associated $1\sigma$ errors are listed in Table~\ref{tab2}. Some formal $\pm 1\sigma$ intervals for the $(1+z)$ exponent include negative values, mostly driven by the highest redshift bin we explore ($z\sim1.5$). Several studies have found that compact galaxy number density starts declining with increasing redshift somewhere between $z\sim1.5$ and $z\sim2$ \citep[e.g.,][]{Barro2013,vanderWel2014}. 

The number density increases strongly in the $0.2<z<0.8$ redshift range only for the sample selected using Definition~4 \citep[ultra-compact galaxies from][]{Cassata2011,Cassata2013}. This sample contains a small number of $0.2<z<0.8$ COSMOS objects (82, see Table~\ref{tab1}), most of which are in the two highest redshift bins ($z\sim0.65$ and $z\sim0.75$) with largest volumes and the densest structures. Thus the trend in number density with redshift is driven by the two high-redshift points with smallest associated uncertainties. A compilation of COSMOS data points and the \citet{Cassata2013} results for ultra-compact galaxies shows a much less pronounced positive trend in galaxy number density with redshift. For all other definitions of galaxy compactness and in both $0.2\leqslant z\leqslant 0.8$ and $0<z\leqslant 1.5$ redshift intervals, constancy of the compact galaxy number density ($\beta\sim0$) is included in the 95\% confidence interval for the best-fit slope of the linear $\log(n)\propto \beta\log(1+z)$ relation. Fits to the data demonstrate that our primary conclusion is independent of definition and sampling; the number density of compact galaxies does not significantly evolve in the $0<z<1.5$ redshift range.    

The lack of evolution in number density with redshift suggests that either 1) compact massive galaxies form in the first $4-5$~Gyrs of cosmic history (down to redshift $z\sim1.5$) and do not evolve in size after that time (static equilibrium), or 2) the size growth of high-redshift compact galaxies is balanced by the emergence of new compact massive galaxies at redshift $z\lesssim1$ \citep[e.g.,][Paper II]{Carollo2013}. To explain the evolution in the average size of massive galaxy with redshift,  many studies have suggested that massive galaxies grow in the redshift range $0<z<1$, mainly through a series of minor mergers \citep[e.g.,][]{Newman2012, Lopez-Sanjuan2012}. Thus these objects are obviously removed from the compact sample and must be replaced by some mechanism. Interestingly, the most massive compact galaxies in the SDSS at $z\sim0$ are recently quenched systems \citep[with ages $\lesssim4$~Gyr, or a formation redshift $z_\mathrm{form}\lesssim0.6$,][]{Trujillo2009,Ferre-Mateu2012}. Furthermore, at least a fraction of compact galaxies at intermediate redshift have $E+A$-type spectra \citep[i.e, show features of young passive systems,][]{Damjanov2014}. Thus compact massive galaxies are forming at intermediate redshift. However, the fraction of newly quenched galaxies in the underlying compact population in this redshift range remains an open question. Possible progenitors of intermediate-redshift compact $E+A$ systems may be compact starbursts evolving in analogy to the formation and evolution of compact massive quiescent galaxies at high redshift (see e.g., \citealt{Toft2014, Barro2014, Williams2014, Stefanon2013, Nelson2014} for observational perspective and  e.g., \citealt{Wellons2015, Zolotov2014} for theoretical consideration).

\section{Conclusion}

We use the COSMOS data to measure the abundance of massive compact quiescent galaxies  at intermediate redshift. We correct the parent spectroscopic sample of quiescent systems  for incompleteness due to both sparse spectroscopic sampling and magnitude-limited size measurements to recover the known mass function for quiescent galaxies in COSMOS. The trend in number density of compact galaxies with redshift in the $0.2\leqslant z \leqslant0.8$ range is, within uncertainties, consistent with constant abundance of these systems at intermediate redshift.

Systematic observational issues have negligible effect on the abundance of compact systems:

\begin {itemize}

\item The definition of compact galaxies does not significantly affect the dependence of their number density on redshift.

\item Within the uncertainties, the number density of compact galaxies does not evolve between $z\sim1.5$ and $z\sim0$.

\item Comparison of a compact sample from BOSS with a similar sample based on COSMOS demonstrates that cosmic variance does not dominate the results.

\end{itemize}    

The evolution of compact galaxies at intermediate redshift produces a population of objects where

\begin{itemize}
   
\item The number density of objects  does not depend strongly on the redshift.

\item The abundance of compact systems with masses $M_\ast>10^{10}\, M_\odot$ at intermediate redshift ranges from $7\times10^{-6}$~Mpc$^{-3}$ to $5\times10^{-4}$~Mpc$^{-3}$, depending on the criteria for compactness.

\item The trend in compact galaxy number density with redshift is insensitive to the adopted minimum stellar mass over the range $10^{10}\, M_\odot<M_\ast\lesssim6\times10^{11}\, M_\odot$.

\item  Thus, if compacts are indeed the seed for large objects, some compacts must form at redshift $z<1$ to replenish the supply.

\end{itemize}      

Analysis of the COSMOS field further suggests that the abundance of intermediate-redshift compact objects may depend on environment (as indicated by \citealt{Valentinuzzi2010a} for the $z\sim0$ regime and by \citealt{Raichoor2012} at high redshift). In Paper II we show that the most massive compact objects in this redshift range lie on the fundamental plane determined by the rest of the quiescent population. This result suggests some commonality of the origin of compact and more extended quiescent objects. Exploration of the links between environment, abundance, and the Fundamental Plane may be further clues to the origin and evolution of compact galaxies.

\acknowledgments

We thank the referee for careful reading of the manuscript and for suggesting the inclusion of Table~\ref{tab2}. ID is supported by the Harvard College Observatory Menzel Fellowship and the Natural Sciences and Engineering Research Council of Canada Postdoctoral Fellowship (NSERC PDF-421224-2012). The Smithsonian Institution supports the research of MJG. HJZ gratefully acknowledges the generous support of the Clay Fellowship.

\end{document}